\documentclass[letterpaper,11pt,fleqn]{article}
\usepackage{jheppub}
\usepackage[latin1]{inputenc}
\usepackage[english]{babel}
\usepackage[numbers,sort&compress]{natbib}
\usepackage{mathabx}
\usepackage{slashed}
\usepackage{yfonts}

\usepackage{graphicx}
\usepackage{dcolumn}
\usepackage{bm}
\usepackage{epstopdf}
\usepackage{mathrsfs}
\usepackage{amssymb,amsfonts,latexsym}
\usepackage{amsmath,bbold}
\allowdisplaybreaks

\long\def\comment#1{ }

\newcommand{\rmd}{{\rm d}}

\makeatletter
  \def\my@tag@font{\normalsize}
  \def\maketag@@@#1{\hbox{\m@th\normalfont\my@tag@font#1}}
  \let\amsmath@eqref\eqref
  \renewcommand\eqref[1]{{\let\my@tag@font\relax\amsmath@eqref{#1}}}
\makeatother

\makeatletter
\def\@fpheader{\relax}
\makeatother

\title{\Large The Gluon-Induced Mueller-Tang Jet Impact Factor at Next-to-Leading Order}

\author[1]{M. Hentschinski,}
\author[2]{J. D. Madrigal Martí­nez,}
\author[3]{B. Murdaca}
\author[4,5]{\& A. Sabio Vera}

\affiliation[1]{Department of Physics, Brookhaven National Laboratory, Upton, NY 11973, USA.}
\affiliation[2]{Institut de Physique Théorique, CEA Saclay, F-91191 Gif-sur-Yvette, France.}
\affiliation[3]{Istituto Nazionale di Fisica Nucleare, Grupo Collegiato di Cosenza, I-87036 Arcavacata di Rende, Cosenza, Italy.}
\affiliation[4]{Instituto de Fí­sica Teórica UAM/CSIC, Nicolás Cabrera 15 \&~Facultad de Ciencias, \\
Universidad Autónoma de Madrid, C.U. Cantoblanco, E-28049 Madrid, Spain.}
\affiliation[5]{CERN, Geneva, Switzerland.}

\abstract{ We complete the computation of the Mueller-Tang jet impact factor at next-to-leading order (NLO) initiated in \cite{quark} and presented in \cite{letter} by computing the real corrections associated to gluons in the initial state making use of Lipatov's effective action. NLO corrections for this effective vertex are an important ingredient for a reliable description of large rapidity gap phenomenology within the BFKL approach.}


\usepackage{graphicx}
\begin{document}
\maketitle

\section{Introduction}

Hard exclusive diffraction processes with large momentum transfer provide an interesting test of the properties of the Balitsky-Fadin-Kuraev-Lipatov (BFKL) pomeron \cite{BFKL}. Within the BFKL all-orders resummation of enhanced rapidity logarithms, this object appears as a bound state of two \emph{reggeized} gluons and the amplitude for pomeron exchange is factorized into a convolution of a universal Green's function and process-dependent impact factors \cite{general}. The BFKL Green's function is known at next-to-leading logarithmic accuracy, both in the forward \cite{forward} and non-forward \cite{nonforward} cases, and a number of impact factors have been also computed at NLO, namely those for colliding partons \cite{imppart1,imppart2,imppart3}, forward jet production \cite{impmn}, forward vector meson production $\gamma^*\to V,\,V=\{\rho^0,\omega,\phi\}$ \cite{impvec} and $\gamma^*\to\gamma^*$ transition \cite{impphot}. These results have allowed for the implementation of next-to-leading BFKL corrections (or at least a subset of them) in the phenomenological description of important observables for the study of QCD in the high-energy limit, like the angular decorrelation of dijets at large rapidity separation \cite{muellernavelet1,muellernavelet2} or the proton structure functions at low values of Bjorken-$x$ \cite{F2}.\\

Next-to-leading corrections to the BFKL Green's function are known to be large and important, since, in particular, they determine the running and normalization scales. What is less expected is that NLO corrections to the impact factors would have such a sizeable effect. However, this turns out to be indeed the case for the available computations of cross-sections including NLO impact factors, namely that for electroproduction of two light vector mesons \cite{electron}, the cross-section and azimuthal decorrelation of Mueller-Navelet jets \cite{muellernavelet1}, and the total inclusive $\gamma^*\gamma^*$ scattering cross-section \cite{chiri}.\\

All the previously referred works probe the BFKL Green's function with zero momentum transfer, $t=0$. On the other hand, BFKL dynamics at finite momentum transfer also has a rich associated phenomenology. In particular, the restriction to the forward case captures the pomeron intercept but misses any information about its slope. Probably the simplest observable allowing to study the $t\ne 0$ BFKL kernel is the cross-section for dijet production with a large rapidity gap, the so-called Mueller-Tang configuration \cite{muellertang}. In this case, the absence of emissions over a large region in rapidity suggests that configurations with color singlet exchange in the $t$-channel, understood in terms of the non-forward BFKL Green's function for $\Delta y_{\rm gap}\gg 1$, should play a major role.\\

\begin{figure}[thb]
  \centering
\parbox{.3\textwidth}{\includegraphics[width = .3\textwidth]{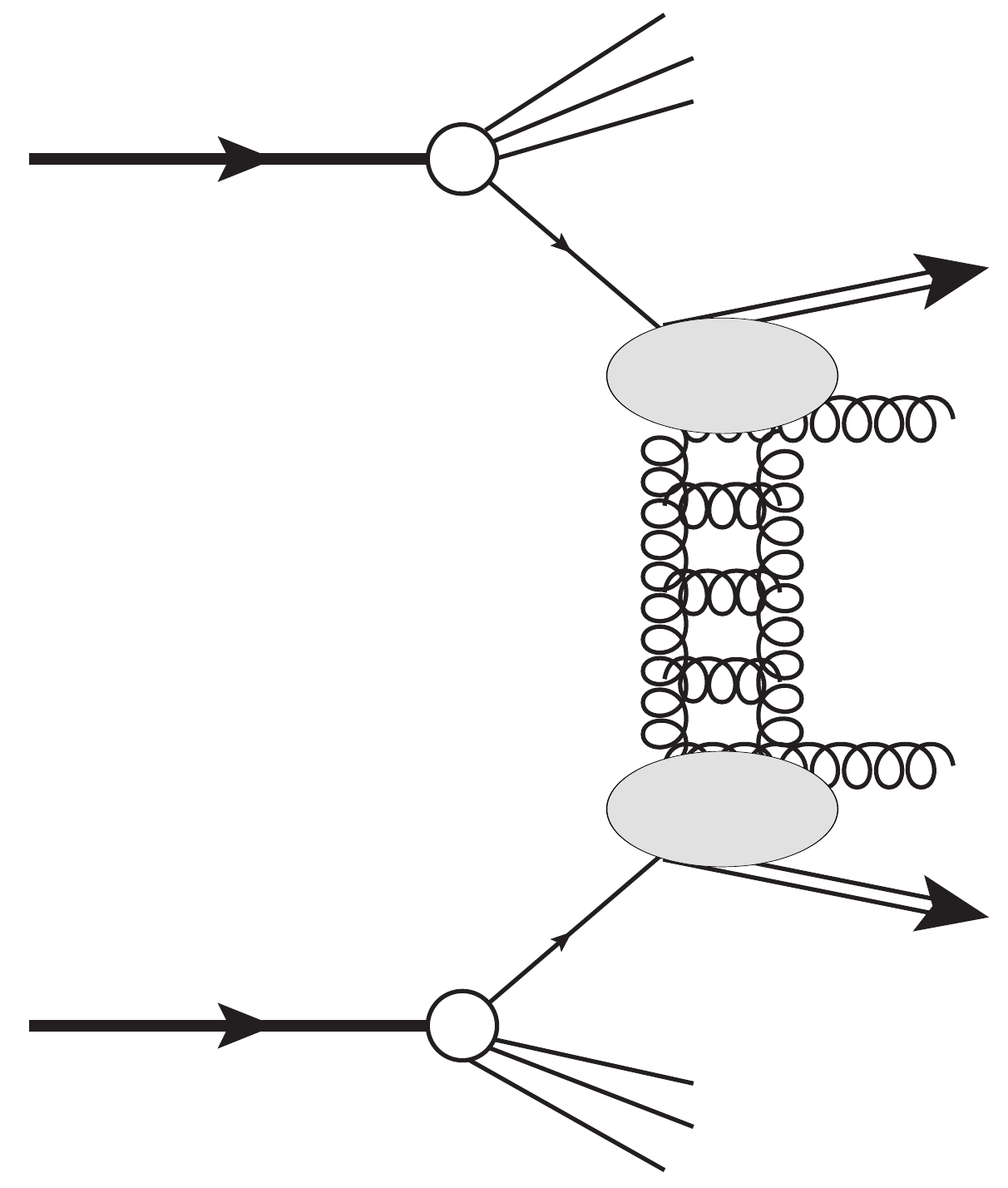}}
\parbox{.3\textwidth}{\includegraphics[width = .3\textwidth]{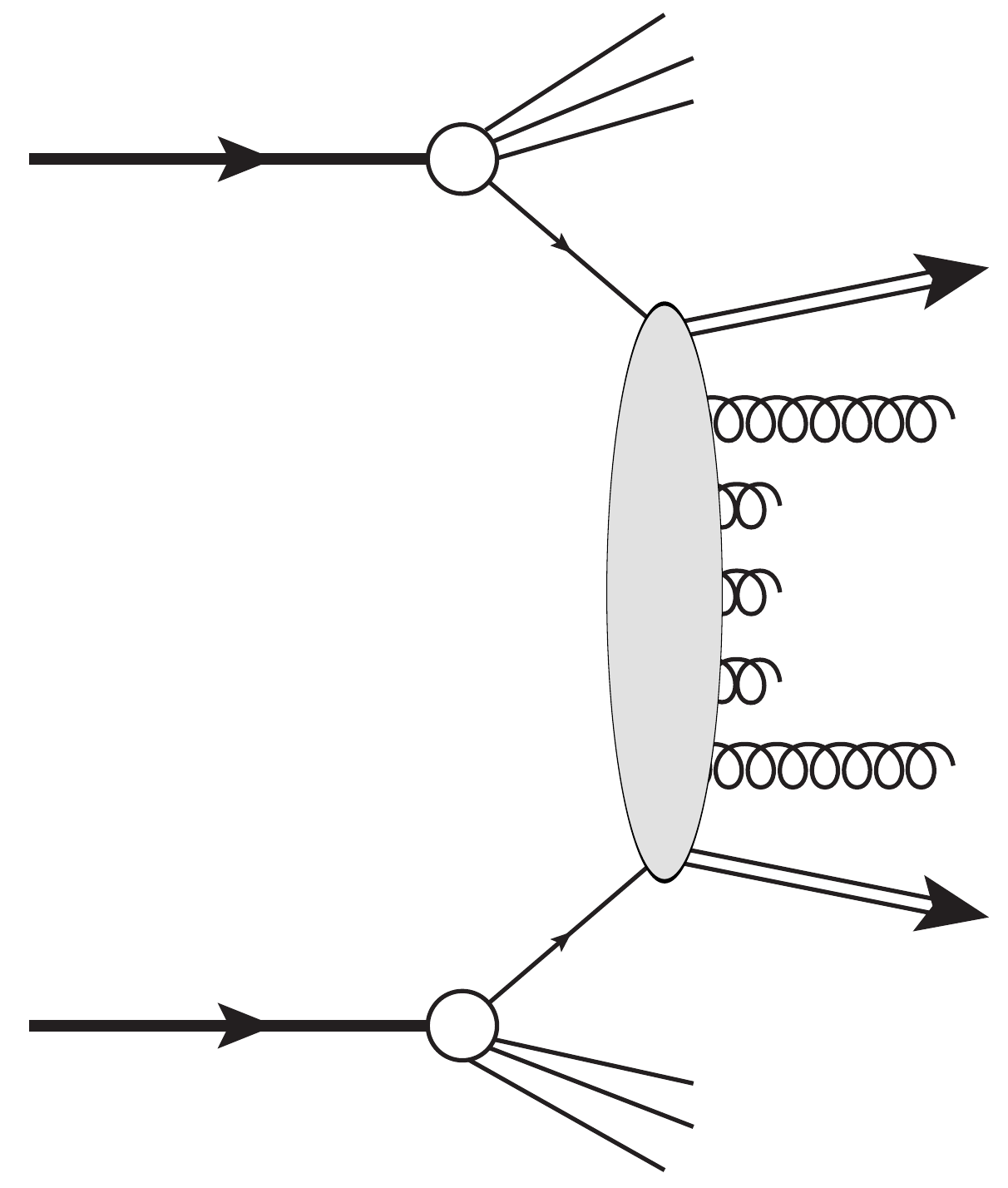}}
\parbox{.3\textwidth}{\includegraphics[width = .3\textwidth]{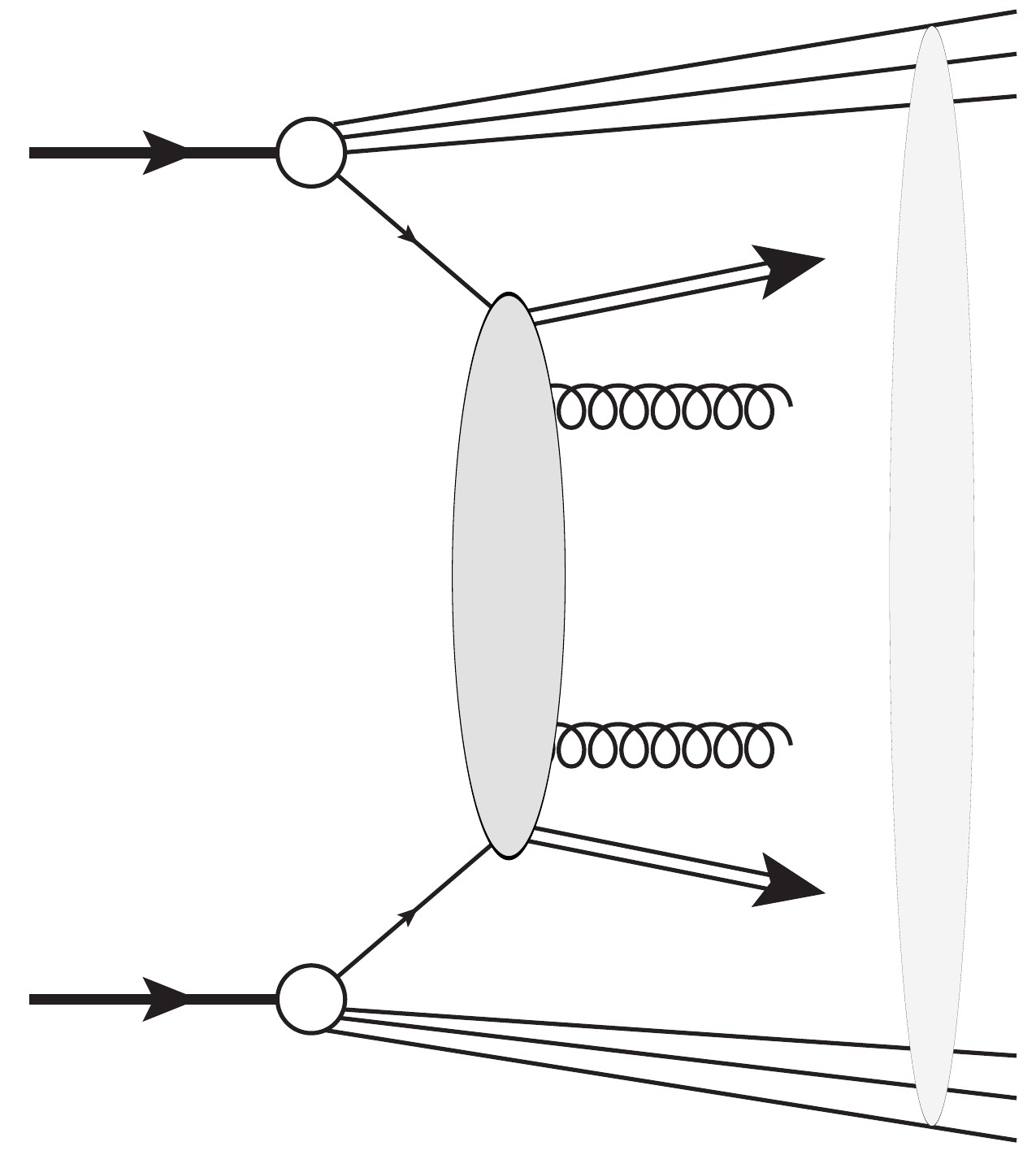}}
\\
\parbox{.3\textwidth}{\center (a)}
\parbox{.3\textwidth}{\center (b)}
\parbox{.3\textwidth}{\center (c)}

  \caption{Different contributions to the Mueller-Tang cross-section: {\em a)} color singlet exchange;     {\em b)} Soft emissions with $p_T$ values smaller than the gap resolution (octet exchange); {\em c)} both contributions can be subject to soft rescattering of the proton remnants which destroy the gap, resulting in a rapidity gap survival factor.}
  \label{fig:different_contributions}
\end{figure}

The original leading-log computation of Mueller and Tang is known not to be able to reproduce the data for rapidity gaps in $p\bar{p}$ collisions obtained at Fermilab/Tevatron \cite{fermilab}. Some improvements have been achieved by the inclusion of the rapidity gap survival factor \cite{cox} (see Fig.~\ref{fig:different_contributions}), constrained kinematics and running coupling corrections \cite{motyka1,motyka2}, and some important NLO corrections at the level of the Green's function \cite{royon}. Notwithstanding, one should also expect large NLO corrections coming from the impact factors, for which only the virtual corrections to elastic parton-parton scattering are currently available \cite{imppart1,imppart2}.\\

In a companion article \cite{quark}, we have computed the missing real emission contribution to the NLO impact factors for quark-initiated jets with color singlet exchange. Here we offer the details of the calculation of the gluon-induced counterpart, thus obtaining the missing ingredient for the full NLO Mueller-Tang impact factor, and hence opening the door to a fully next-to-leading description of events with large rapidity gaps at high momentum transfer. In both computations, we use Lipatov's effective action \cite{effective}, whose application has been recently extended beyond the calculation of tree-level scattering amplitudes in quasi-multi-Regge kinematics by providing suitable regularization and subtraction prescriptions \cite{ours}.\\

In Section \ref{2} we introduce our notation, while we refer to \cite{effective,ours} for a deeper introduction to Lipatov's action and its application to amplitudes at loop level. Afterwards, in Section \ref{3} we address the computation of the gluon-initiated real-emission corrections. Section \ref{4} is devoted to the jet definition and the NLO description of Mueller-Tang jets within collinear factorization. We check that after introducing the jet definition and integrating over the real particle phase space, those soft and collinear singularities not reabsorbed in the renormalization of the coupling and of the parton distribution functions cancel among virtual and real corrections. Finally, we present some general remarks. An Appendix deals with the explicit results for the inclusive (perturbative) pomeron-gluon impact factor.

\section{Mueller-Tang Jets at Parton Level and the High-Energy Effective Action}\label{2}

\begin{figure}[htb]
    \label{fig:subfigures}
   \centering
   \parbox{.7cm}{\includegraphics[height = 1.8cm]{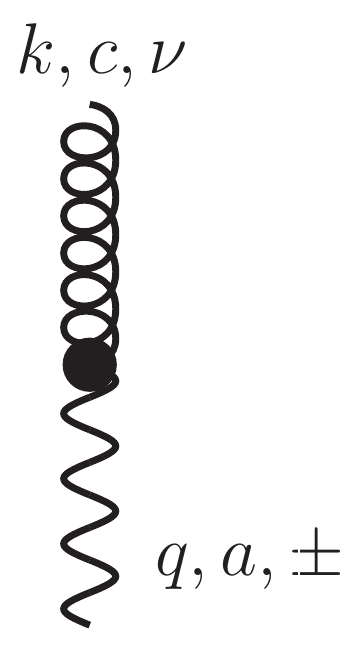}} $=  \displaystyle 
   \begin{array}[h]{ll}
    \\  \\ - i{\bm q}^2 \delta^{a c} (n^\pm)^\nu,  \\ \\  \qquad   k^\pm = 0.
   \end{array}  $ 
 \parbox{1.2cm}{ \includegraphics[height = 1.8cm]{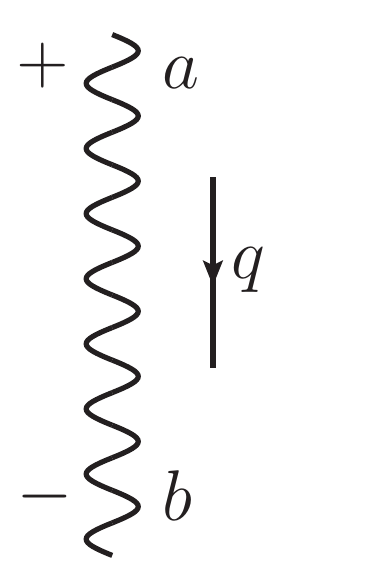}}  $=  \displaystyle    \begin{array}[h]{ll}
    \delta^{ab} \frac{ i/2}{{\bm q}^2} \end{array}$ \\
 \parbox{1.7cm}{\includegraphics[height = 1.8cm]{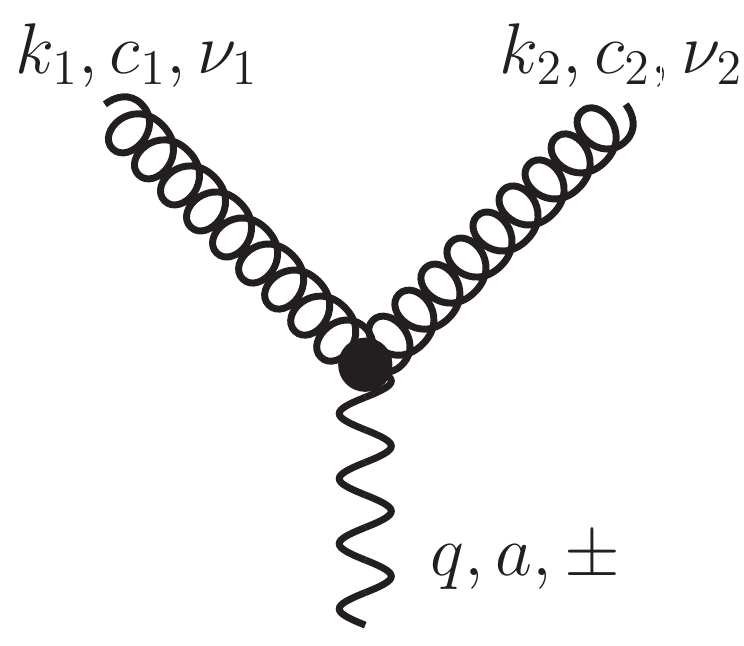}} $ \displaystyle  =  \begin{array}[h]{ll}  \\ \\ g f^{c_1 c_2 a} \frac{{\bm q}^2}{k_1^\pm}   (n^\pm)^{\nu_1} (n^\pm)^{\nu_2},  \quad  k_1^\pm  + k_2^\pm  = 0
 \end{array}$\\
  \parbox{2.8cm}{\includegraphics[height = 1.8cm]{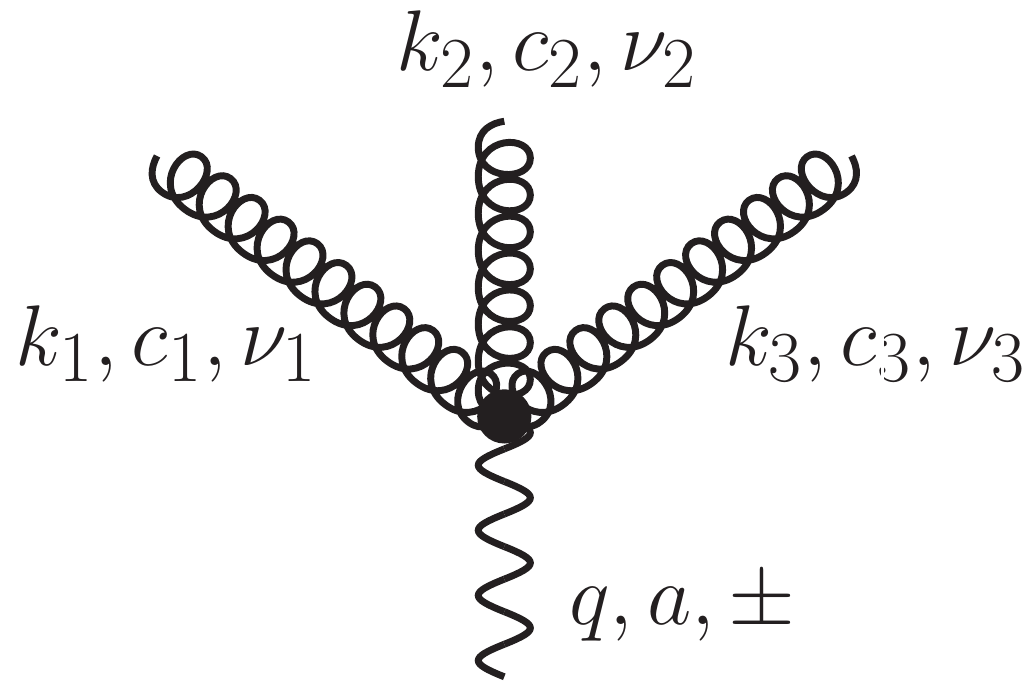}} $ \displaystyle  =  \begin{array}[h]{ll}  \\ \\ ig^2\bm{q}^2 \left(\frac{f^{c_3c_2c}f^{c_1ca}}{k_3^\pm k_1^\pm}+\frac{f^{c_3c_1c}f^{c_2ca}}{k_3^\pm k_2^\pm}\right)   (n^\pm)^{\nu_1} (n^\pm)^{\nu_2}(n^\pm)^{\nu_3},  \\ \\ \quad  k_1^\pm  + k_2^\pm + k_3^\pm = 0
 \end{array}$
 \caption{\small Feynman rules for the lowest-order effective vertices of the effective action \cite{antonov}. Wavy lines denote reggeized gluons and curly lines, gluons. Pole prescriptions for the light-cone denominators are discussed in \cite{hentschinski}.}
\label{fig:3}
\end{figure}

The process under study, for the sake of concreteness, will be dijet production in a $pp$ collision,
\begin{equation}\label{defs}
p(p_A)+p(p_B)\to J_1(p_{J,1})+J_2(p_{J,2})+\text{gap},
\end{equation}
where the two jets are tagged at a large rapidity separation which includes a large region $\Delta y_{\rm gap}$ devoid of hadronic activity. We focus on color singlet exchange in the $t$-channel. The presence of such a large rapidity separation invokes the use of high-energy factorization for scattering amplitudes in multi-Regge kinematics, which is conveniently embodied in the following effective action put forward by Lipatov \cite{effective}
\begin{equation}
S_{\rm eff}=S_{\rm QCD}+S_{\rm ind};\qquad S_{\rm ind}=\int {\rm d}^4 x\,{\rm Tr}[(W_-[v(x)]-A_-(x))\partial_\perp^2 A_+(x)+\{+\leftrightarrow -\}],
\end{equation}
where $v_\mu=-iT^av_\mu^a(x)$ is the gluon field, and $A_\pm(x)=-iT^aA^a_\pm(x)$ is the {\em reggeon} field, introduced as a new degree of freedom, which mediates any interaction between (clusters of) particles highly separated in rapidity. On the other hand, local-in-rapidity interactions between reggeons and gluons are mediated by the Wilson line couplings $W_\pm [v(x)]=-\frac{1}{g}\partial_\pm {\cal P}\exp \left\{-\frac{g}{2}\int_{-\infty}^{x^\mp} dz^\pm v_\pm (z)\right\}$. The reggeon field satisfies the kinematic constraint $\partial_\pm A_\mp (x)$ $=0$, manifest in the momentum space Feynman rules of Fig. \ref{fig:3}. We have introduced the Sudakov decomposition
\begin{equation}
k=k^+\frac{n^-}{2}+k^-\frac{n^+}{2}+\bm{k};\qquad
n^\pm=2p_{A,B}/\sqrt{s},\quad s=2p_A\cdot p_B,
\end{equation}
where $p_{A,B}$ are the momenta of the colliding hadrons. For the process \eqref{defs}, we have
\begin{align}
  \label{eq:SudaAMP}
  p_A & = p_A^+ \frac{n^-}{2}, & 
p_{J,1} & = \sqrt{{\bm k}_{J,1}^2}  \left(  e^{y_{J,1}}\frac{n^-}{2} +   e^{-y_{J,1}} \frac{n^+}{2} \right) + {\bm k}_{J,1}. \notag \\
  p_B & = p_B^- \frac{n^+}{2}, & p_{J,2} & =  \sqrt{{\bm k}_{J,2}^2}  \left(  e^{y_{J,2}}\frac{n^-}{2} +   e^{-y_{J,2}} \frac{n^+}{2} \right) + {\bm k}_{J,2}.
\end{align}

Here $(\bm{k}_{J,i},y_{J,i}),\,i=1,2$, are the transverse momenta and rapidity of the jets. At the partonic level, we will be concerned with the process
\begin{equation}\label{partonic}
g(p_a)+g(p_b)\to g(p_1)+g(p_2),
\end{equation}
mediated by color-singlet in the $t$-channel, which requires the exchange of (at least) two reggeons. The cross-section for \eqref{partonic}, at NLO, will read
\begin{equation}\label{kreutz}
d\hat{\sigma}_{ab}=\left[\int\frac{{\rm d}^{2+2\epsilon}\bm{l}_1}{\pi^{1+\epsilon}}\frac{1}{\bm{l}_1^2(\bm{k}-\bm{l}_1)^2}\int\frac{{\rm d}^{2+2\epsilon}\bm{l}_2}{\pi^{1+\epsilon}}\frac{1}{\bm{l}_2^2(\bm{k}-\bm{l}_2)^2}h_{g,{\rm a}}h_{g,{\rm b}}\right]d^{2+2\epsilon}\bm{k},\quad h_g=h_g^{(0)}+h_g^{(1)},
\end{equation}
for bare reggeon exchange, while after the resummation of $\Delta y_{\rm gap}\sim\ln(\hat{s}/s_0)$ enhanced terms to all orders in $\alpha_s$, we get\footnote{Notice that the Green's function is related to the {\em imaginary part} of the amplitude rather than the amplitude itself, which is the object needed to compute the exclusive cross-section with finite momentum transfer. At leading order, the amplitude for singlet exchange is purely imaginary, while this is no longer true at next-to-leading order, since one has a signature factor
\begin{equation}
G(\bm{l},\bm{l}',\bm{k},s/s_0)=\int_{\delta-i\infty}^{\delta+i\infty}\frac{{\rm d}\omega}{2\pi i}\left(\frac{s}{s_0}\right)^{\omega+1} G(\bm{l},\bm{l}',\bm{k},\omega)\to \int_{\delta-i\infty}^{\delta+i\infty}\frac{{\rm d}\omega}{2\pi i}\left(\frac{s}{s_0}\right)^{\omega+1}\frac{1-e^{-i\pi\omega}}{\sin \pi\omega}G(\bm{l},\bm{l}',\bm{k},\omega).
\end{equation}
However, this effect of order ${\cal O}(\alpha_s)$ will cancel among the amplitude and its complex conjugate, and \eqref{resum} remains valid at NLO. Our normalization of the BFKL equation is the one by Forshaw and Ross \cite{general}.}
\begin{equation}\label{resum}
\rmd\hat{\sigma}_{ab}^{\text{res}}=\left[\int\frac{{\rm d}^2\bm{l}_1}{\pi}\int{\rm d}^2\bm{l}_1'\,G(\bm{l}_1,\bm{l}_1',\bm{k},\hat{s}/s_0)\int\frac{{\rm d}^2\bm{l}_2}{\pi}\int{\rm d}^2\bm{l}_2'\,G(\bm{l}_2,\bm{l}_2',\bm{k},\hat{s}/s_0)h_{g,{\rm a}}h_{g,{\rm b}}\right]\rmd^2\bm{k},
\end{equation}
in terms of the non-forward BFKL Green's function $G(\bm{l},\bm{l}',\bm{q},s/s_0)$, with $s_0$ the reggeization scale, which parametrizes the scale uncertainty associated to the resummation. We assume that, in the asymptotic limit $s\to\infty$, the Green's function regulates the infrared divergence associated to the transverse momentum integral, as it occurs at leading log accuracy \cite{motyka2}.\\

The whole dependence of the impact factors on $s_0$ is contained in the virtual corrections to the process \eqref{partonic}, already computed in \cite{imppart2}, where it was checked that the $s_0$ dependence of the cross-section cancels when the Green's function is truncated to NLO. Apart from these contributions, we will need to determine the amplitude for the processes\footnote{At the same order in perturbation theory one could have the exchange of three reggeized gluons in an overall singlet state (odderon) interfering with pomeron exchange. However, the amplitude for pomeron exchange is imaginary while that for odderon exchange is real, and therefore the possible interference vanishes.}
\begin{subequations}\label{processes}
\begin{alignat}{2}
g(p_a)+g(p_b)&\to g(p)+g(p_2)+g(q),\label{p1}\\
g(p_a)+g(p_b)&\to q(p)+g(p_2)+\bar{q}(q),\label{p2}
\end{alignat}
\end{subequations}
with color singlet exchange in one of the $t$-channels $t_1=(p_a-p_1)^2$ and $t_2=(p_b-p_2)^2$. This is the goal of the next section.

\begin{figure}[th]
  \centering
 \parbox{4cm}{\center \includegraphics[height = 2.2cm]{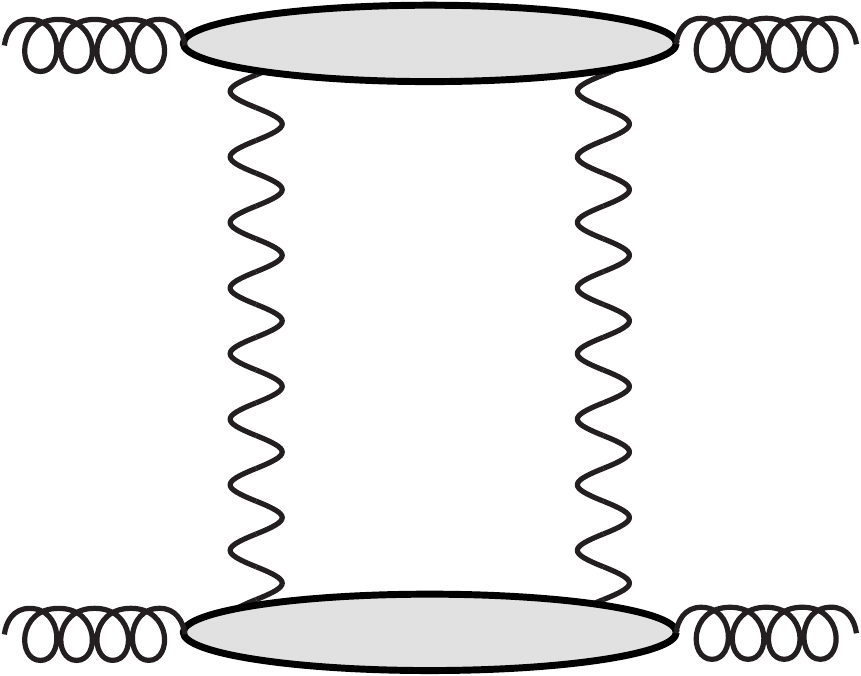}}
\parbox{.7\textwidth}{\center\includegraphics[width = 2.5cm]{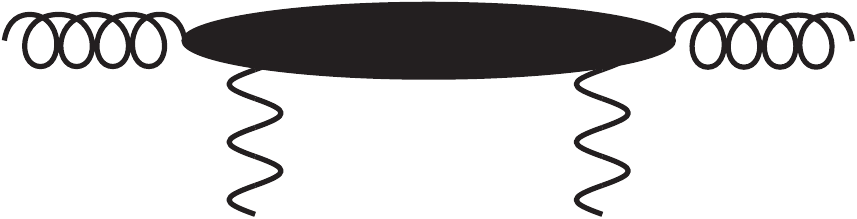} $=$  
\includegraphics[width = 2.5cm]{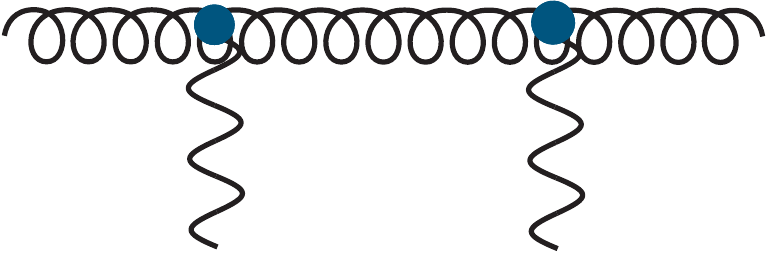} + \includegraphics[width = 2.5cm]{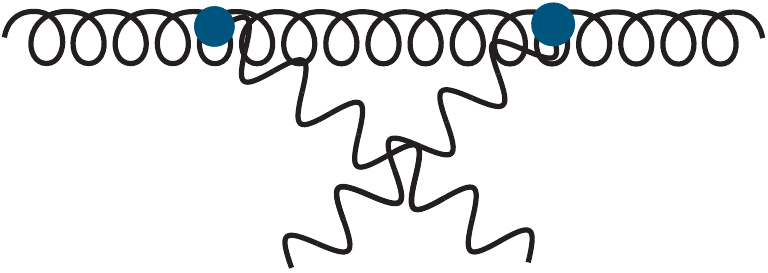};\quad \includegraphics[width = 2.5cm]{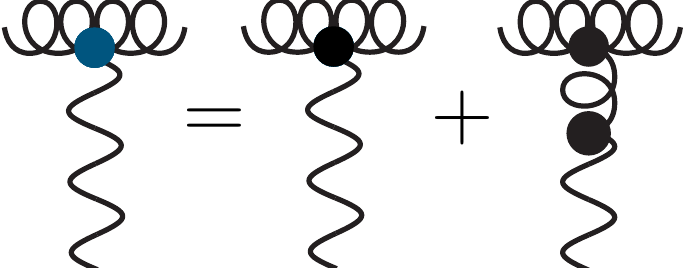}.}  
\parbox{4cm}{\center (a)}\parbox{.7\textwidth}{\center (b)}
  \caption{{\em a)} The leading log amplitude for gluon induced jets in the high-energy approximation. The state of two reggeized gluons in the $t$-channel is projected onto the color singlet; {\em b)} Leading order diagrams which describe within the effective action the coupling of the gluon to the two reggeons.}
  \label{fig:schematic}
\end{figure}

\section{Real NLO Corrections to the Impact Factors}\label{3}

\subsection{The Gluon-Initiated Mueller-Tang Cross-Section at LO}\label{31}
Before discussing the computation of NLO real corrections, let us review how the leading log result is obtained from Lipatov's action. The diagrams of interest are shown in Fig. \ref{fig:schematic}.
The $t$-channel projector onto the color singlet is
\begin{equation}\label{projector}
{\cal P}^{ab,a'b'}=P^{ab}P^{a'b'},\quad P^{ab}=\frac{\delta^{ab}}{\sqrt{N_c^2-1}}.
\end{equation}
Then, at leading order, the scattering amplitude in the high-energy limit for the process \eqref{partonic}, projected into the color singlet, reads
\begin{equation}\label{effenberg}
i{\cal M}^{(0)}_{g_ag_b\to g_1g_2}=\frac{1}{2\cdot 2!}\int\frac{{\rm d}l^+ {\rm d}l^-}{(2\pi)^2}\int\frac{{\rm d}^{2+2\epsilon}\bm{l}}{(2\pi)^{2+2\epsilon}}i\tilde{\cal M}^{abde}_{g2r_+^*\to g}\cdot i\tilde{\cal M}^{a'b'd'e'}_{g2r^*_-\to g}{\cal P}^{de,d'e'}\frac{(i/2)^2}{\bm{l}^2(\bm{l}-\bm{k})^2},
\end{equation}
where the 2! denominator corrects the overcounting from exchanging the reggeons in both the upper and lower sides of the amplitude. The Sudakov decomposition of the sub-process $g(p_a)+r^*_+(l_1)+r_+^*(k-l_1)\to g(p)$, with $r^*$ denoting the virtual reggeons, is
 \begin{align}
   \label{eq:sudaBORN}
   p_a &= p_a^+ \frac{n^-}{2} &  p &=  p_a^+\frac{n^-}{2} + k^- \frac{n^+}{2} + {\bm k} \notag \\
 l_1 & = l_1^- \frac{n^+}{2} + {\bm l}_1 & k&=  k^-\frac{n^+}{2} + {\bm k},
 \end{align}
with
\begin{equation}
i\tilde{\cal M}^{abde}_{g2r_+^*\to g}P^{de}=\frac{4ig^2p_a^+N_c\,}{\sqrt{N_c^2-1}}\delta_{ab}\,\varepsilon(p_a)\cdot\varepsilon^*(p)\left[\frac{1}{l^--\frac{\bm{l}^2}{p_a^+}+i0}+\frac{1}{(k-l)^--\frac{(\bm{k}-\bm{l})^2}{p_a^+}+i0}\right].
\end{equation}
High-energy factorization implies that the entire dependence on the longitudinal loop momenta $l^-$ and $l^+$ is contained in the $gr^*r^*\to g$ amplitudes, even when considering higher radiative corrections to the impact factors. This allows us to express the amplitude \eqref{effenberg} in terms of a unique transverse loop integral:
\begin{equation}\label{linke}
i{\cal M}^{(0)}_{g_a g_b\to g_1 g_2}=\int\frac{{\rm d}^{2+2\epsilon}\bm{l}}{(2\pi)^{2+2\epsilon}}\phi_{gg,{\rm a}}\phi_{gg,{\rm b}}\frac{1}{\bm{l}^2(\bm{k}-\bm{l})^2},
\end{equation}
with
\begin{equation}
i\phi_{gg,{\rm a}}=\int\frac{{\rm d}l^-}{8\pi}i\tilde{\cal M}^{abde}_{gr^*r^*\to g}P^{de}=\frac{N_c}{\sqrt{N_c^2-1}}g^2p_a^+\delta^{ab}\varepsilon_\lambda(p_a)\varepsilon_\mu^*(p)g^{\lambda\mu}.
\end{equation}
The choice of gauge for the polarization vectors is
\begin{equation}
\varepsilon_{\lambda}(q,n^+)\cdot q=\varepsilon_{\lambda}(q,n^+)\cdot n^+=0 \Longrightarrow \varepsilon^\mu_{\lambda}(q,n^+)=\frac{\bm{\varepsilon}_{\lambda}\cdot\bm{q}}{q^+}(n^+)^\mu+\bm{\varepsilon}^\mu_{\lambda}.
\end{equation}
To extract the leading order impact factor, we consider the general definition
\begin{equation}
d\hat{\sigma}_{ab}=\frac{1}{\Phi}\overline{|{\cal M}^{(0)}_{g_ag_b\to g_1g_2}|^2}d\Pi_2;\quad \Phi\stackrel{s\to\infty}{\simeq} 2s,
\end{equation}
with the differential phase space
\begin{equation}
\begin{aligned}
d\Pi_2&=\iint \frac{{\rm d}^dp_1}{(2\pi)^{d-1}}\frac{{\rm d}^dp_2}{(2\pi)^{d-1}}\delta (p_1^2)\delta(p_2^2)(2\pi)^d\delta^{(d)}(p_1+p_2-p_a-p_b)=\int \frac{{\rm d}^dp_1}{(2\pi)^{2+2\epsilon}}\delta(p_1^2)\\&\times\delta ((p_a+p_b-p_1)^2)=\int \frac{{\rm d}^{\rm d}k}{(2\pi)^{2+2\epsilon}}\delta((p_a-k)^2)\delta((p_b-k)^2)=\frac{1}{2p_a^+p_b^-}\int \frac{\rmd^{2+2\epsilon}\bm{k}}{(2\pi)^{2+2\epsilon}},
\end{aligned}
\end{equation}
where we have used that $(p_a+k)^2=p_a^+k^-,\,(p_b-k)^2=-p_b^-k^+,$ and ${\rm d}^{\rm d}k=\frac{1}{2}{\rm d}k^+{\rm d}k^-{\rm d}^{d-2}\bm{k}$. From \eqref{effenberg} and \eqref{linke}, we get the squared amplitude, summed over final color and polarization indices, and averaged over initial ones, {\it i.e.} 
\begin{equation}
\begin{aligned}
&\overline{|{\cal M}^{(0)}_{g_ag_b\to g_1g_2}|^2}=\int\frac{{\rm d}^{2+2\epsilon}\bm{l}_1}{(2\pi)^{2+2\epsilon}}\frac{1}{\bm{l}_1^2(\bm{k}-\bm{l}_1)^2}\int\frac{{\rm d}^{2+2\epsilon}\bm{l}_2}{(2\pi)^{2+2\epsilon}}\frac{1}{\bm{l}_2^2(\bm{k}-\bm{l}_2)^2}\overline{|\phi_{gg,{\rm a}}|^2}\overline{|\phi_{gg,{\rm b}}|^2};\\
\overline{|\phi_{gg,{\rm a}}|^2}&=\frac{1}{2(N_c^2-1)}\sum_{a,b}\delta^{ab}\frac{N_c^2}{N_c^2-1}g^4(p_a^+)^2\bigg\{g^{\lambda\mu}g^{\lambda'\mu'}\left[-g_{\lambda\lambda'}+\frac{p_{a\lambda}(n^+)_{\lambda'}+(n^+)_\lambda p_{a\lambda'}}{p_a^+}\right]\\
&\times\left[-g_{\mu\mu'}+\frac{p_\mu(n^+)_{\mu'}+(n^+)_\mu p_{\mu'}}{p^+}\right]\bigg\}=(1+\epsilon)\frac{N_c^2}{N_c^2-1}g^4(p_a^+)^2.
\end{aligned}
\end{equation}
This yields the expression for the cross-section
\begin{equation}
\begin{aligned}
d\hat{\sigma}_{ab}&=\frac{1}{(2p_a^+p_b^-)^2}\left[\frac{(1+\epsilon)N_c^2}{N_c^2-1}\frac{g^4}{(8\pi^2)^{1+\epsilon}}\right]^2(p_a^+)^2(p_b^-)^2\\&\times\left[\int\frac{{\rm d}^{2+2\epsilon}\bm{l}_1}{\pi^{1+\epsilon}}\frac{1}{\bm{l}_1^2(\bm{k}-\bm{l}_1)^2}\right]\left[\int\frac{{\rm d}^{2+2\epsilon}\bm{l}_2}{\pi^{1+\epsilon}}\frac{1}{\bm{l}_2^2(\bm{k}-\bm{l}_2)^2}\right]d^{2+2\epsilon}\bm{k},
\end{aligned}
\end{equation}
and compared to \eqref{kreutz}, we get
\begin{equation}\label{krauss}
h_g^{(0)}=\frac{\overline{|\phi_{gg,{\rm a}}|^2}}{2(8\pi^2)^{1+\epsilon}(p_a^+)^2}=h^{(0)}(1+\epsilon)C_a^2,\qquad h^{(0)}=\frac{\alpha_{s,\epsilon}^22^\epsilon}{\mu^{4\epsilon}\Gamma^2(1-\epsilon)(N_c^2-1)},
\end{equation}
where we have introduced the strong coupling in $\overline{\rm MS}$ scheme in $d=4+2\epsilon$ dimensions:
\begin{equation}
\alpha_{s,\epsilon}=\frac{g^2\mu^{2\epsilon}\Gamma(1-\epsilon)}{(4\pi)^{1+\epsilon}}.
\end{equation}
\subsection{The Real NLO Corrections to the Impact Factor}

\begin{figure}[thb]
  \centering
  \parbox{3cm}{\center \includegraphics[height=2.5cm]{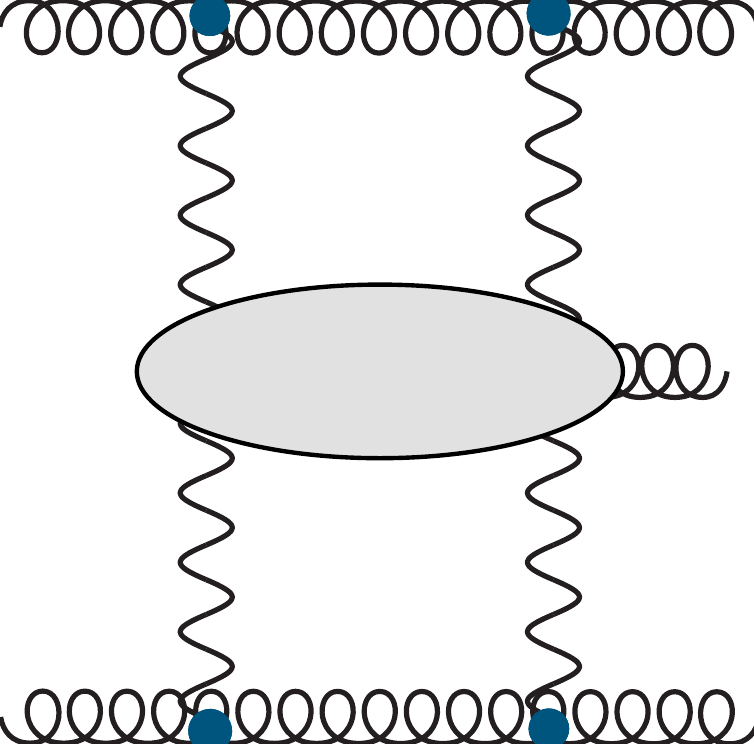}} 
  \parbox{3cm}{\center \includegraphics[height=2.5cm]{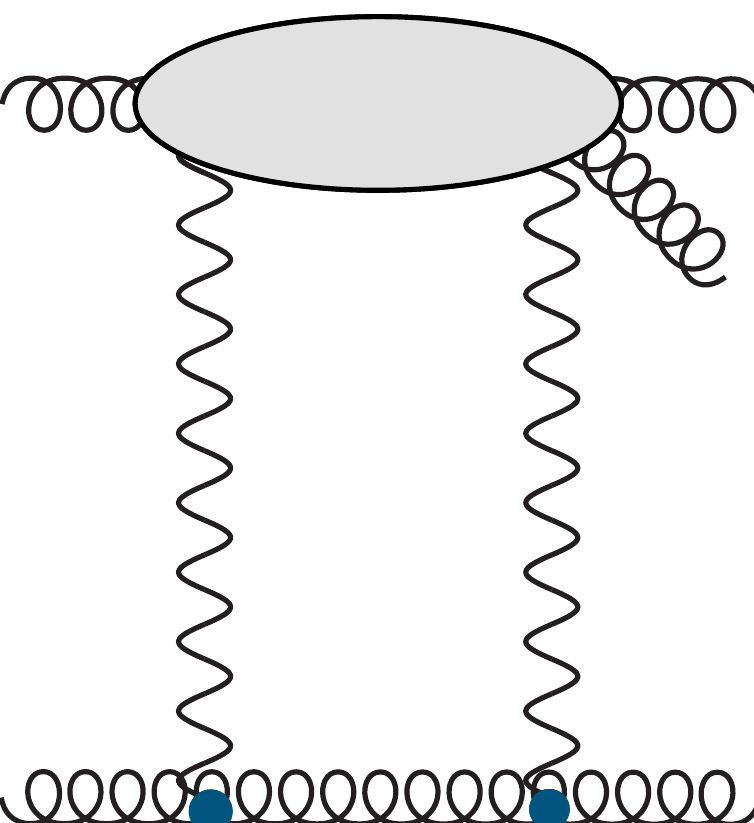}}
  \parbox{3cm}{\center \includegraphics[height=2.5cm]{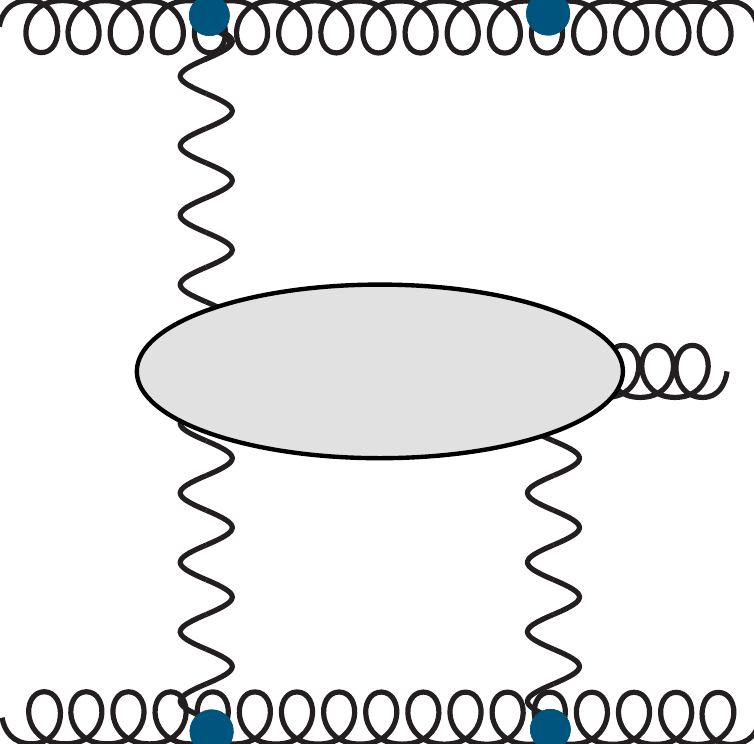}}
  \parbox{3cm}{\center \includegraphics[height=2.5cm]{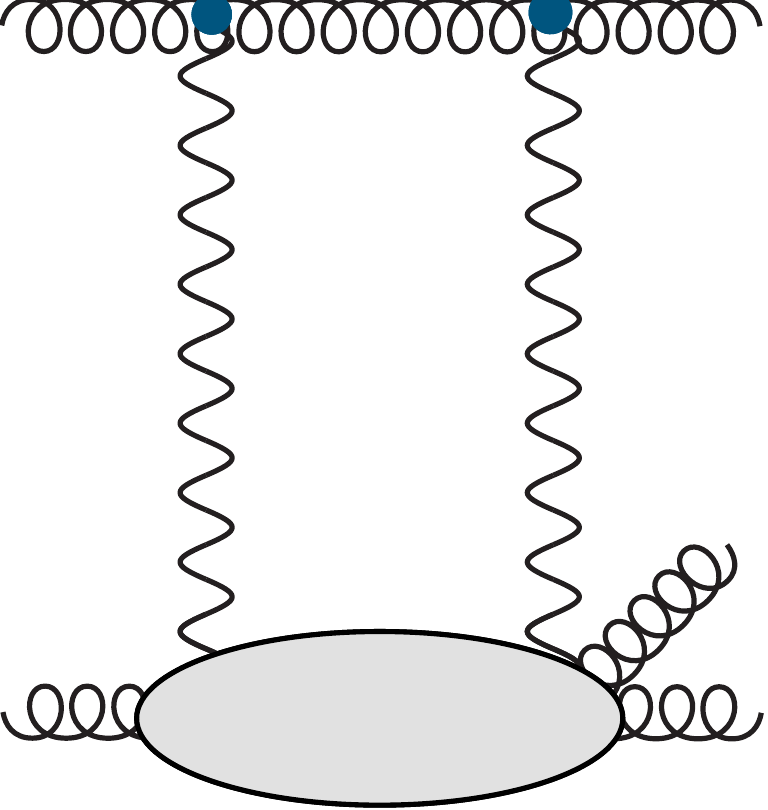}}      
  \parbox{3cm}{\center \includegraphics[height=2.5cm]{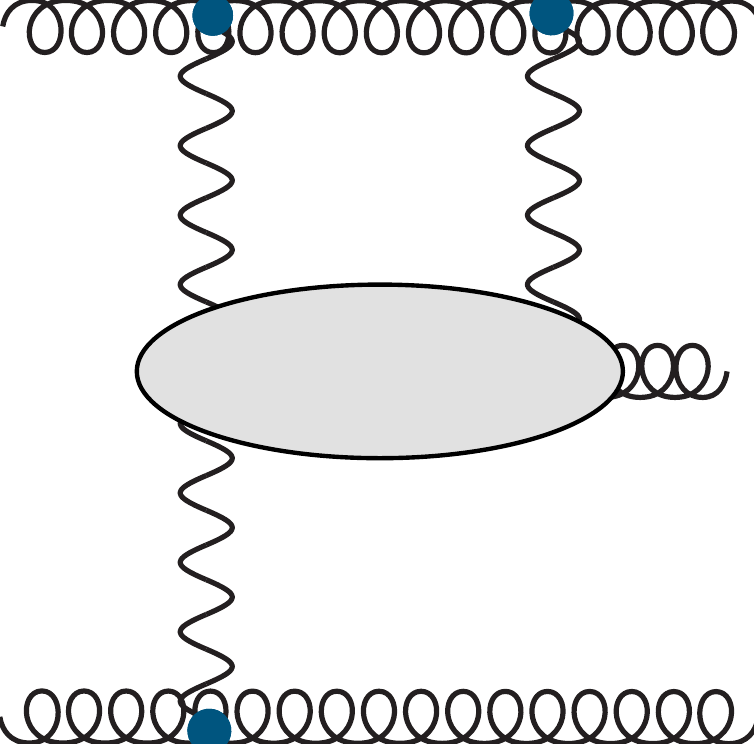}} 
  \parbox{3cm}{\center (a) }
  \parbox{3cm}{\center (b) }
  \parbox{3cm}{\center (c) }
  \parbox{3cm}{\center (d) }
  \parbox{3cm}{\center (e) }

  \caption{Real NLO corrections with $gg$ final state.}
  \label{fig:realNLO}
\end{figure}

\begin{figure}[htb]
  \centering
 \parbox{.3\textwidth}{ \center \includegraphics[width = .2\textwidth]{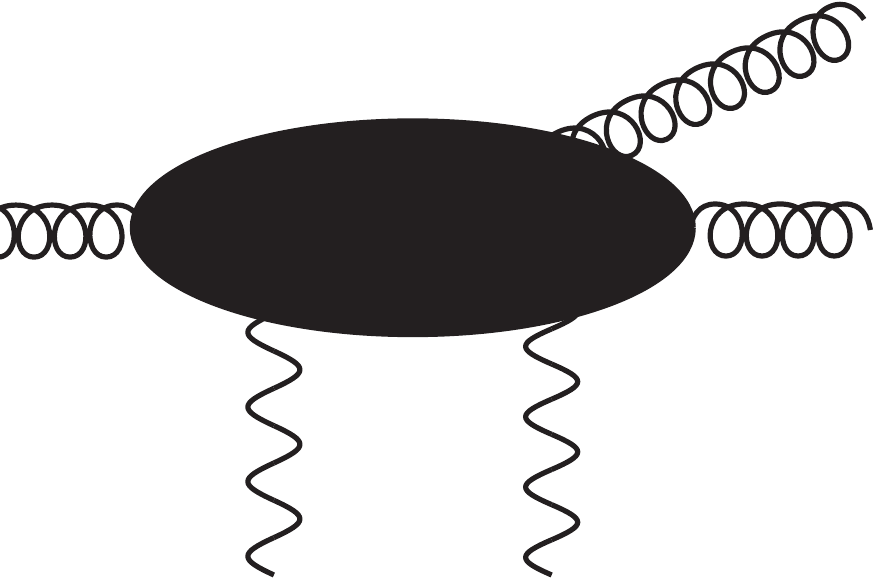} } 
 \parbox{.3\textwidth}{\center \includegraphics[width = .2\textwidth]{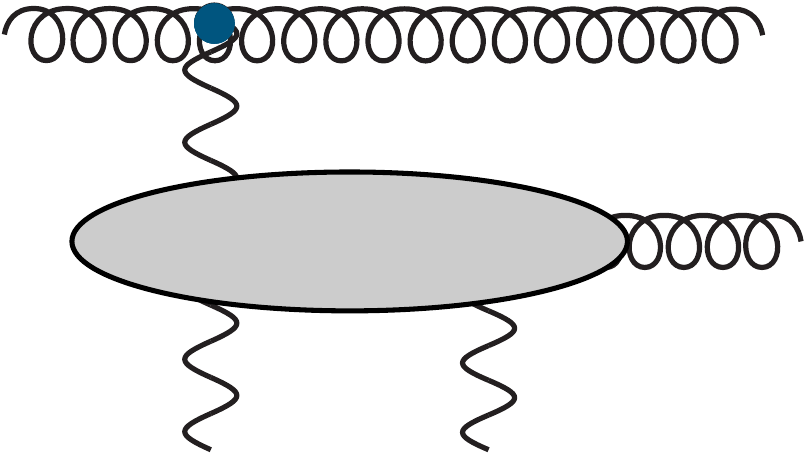}}
\parbox{.3\textwidth}{\center  \includegraphics[width = .145\textwidth]{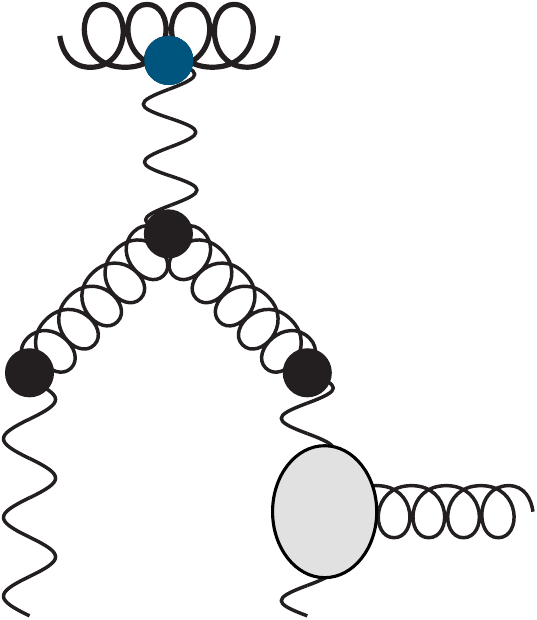}}

 \parbox{.3\textwidth}{ \center (a)}  \parbox{.3\textwidth}{ \center (b)}  \parbox{.3\textwidth}{ \center (c)}
  \caption{Different reggeon diagrams contributing to the real corrections to the Mueller-Tang impact factor: {\em a)} Quasielastic and {\em b)} Central production diagram; {\em c)} Diagram with a $r^*\to 2 r^*$ splitting. The grey blob denotes Lipatov's effective vertex. Those contributions can be seen to vanish identically after integration over $l^-$, if the Hermiticity of the effective action is respected by using the pole prescription discussed in \cite{hentschinski}. It is understood that no internal reggeon lines appear inside the blobs.}
  \label{fig:regg}
\end{figure}

In Fig. \ref{fig:realNLO}, we have schematically shown the different NLO real corrections to the process \eqref{p1}; similar diagrams can be written for the quasielastic corrections (b) and (d) for the $q\bar{q}$ final state \eqref{p2}. The different contributions can be sorted out into two pieces: those with reggeon exchange in both $t$-channels (Fig. \ref{fig:realNLO}, (a), (c), (e)), corresponding to gluon emission at central rapidities, and those where the additional gluon is emitted in the fragmentation region of one of the gluons (quasielastic contribution, Fig. \ref{fig:realNLO}, (b), (d)). As discussed at length in \cite{quark}, only quasielastic contributions will be relevant for the impact factor for jet production with a rapidity gap.
In any case, the central production amplitude (Fig. \ref{fig:regg}, (b)), appearing in diagrams (c) and (e) of Fig. \ref{fig:realNLO}, provides a useful check of our computation, already exploited in \cite{quark}, since the limit of the quasielastic amplitude (Fig. \ref{fig:regg} (a)), 
\begin{equation}\label{imass}
\hat{M}_X^2=\{s_{gg},\,s_{q\bar{q}}\}=\frac{(z\bm{p}+(1-z)\bm{q})^2}{z(1-z)}\to\infty,
\end{equation}
must coincide with the central production amplitude.

\subsection{Computation of the Quasielastic Corrections}\label{33}
Let us consider now the processes $g(p_a)+r^*(l)+r^*(k-l)\to \{g(p)+g(q); q(p)+\bar{q}(q)\}$\footnote{Throughout the text, the momenta of initial particles are considered incoming, while those of final particles are outgoing.}, with the following Sudakov decomposition of external momenta
\begin{align}
  \label{eq:Suda_central}
 p_a & = p_a^+ \frac{n^-}{2}, \qquad     \qquad  \qquad 
   k  = k^- \frac{n^+}{2} + {\bm k}, &
  l & =  l^- \frac{n^+}{2} + {\bm l}, \notag  \\ p & = (1-z)p_a^+ \frac{n^-}{2} + \frac{{\bm p}^2}{(1-z) p_a^+} \frac{n^+}{2} +  {\bm p}, &  
 q & = z p_a^+  \frac{n^-}{2} + \frac{{\bm q}^2}{z p_a^+} \frac{n^+}{2} + {\bm q} \,.
\end{align}
The associated Mandelstam invariants are
\begin{equation}\label{mandelstam}
\begin{aligned}
s&=(p_a+k)^2=(p+q)^2=\frac{(\bm{q}-z\bm{k})^2}{z(1-z)};\quad t=(p_a-p)^2=(q-k)^2=-\frac{(\bm{k}-\bm{q})^2}{1-z};\\
\,u&=(p_a-q)^2=(p-k)^2=-\frac{\bm{q}^2}{z};\qquad\quad\,\, s+t+u=-\bm{k}^2.
\end{aligned}
\end{equation}
\begin{figure}[htp]
\centering
\includegraphics[scale=0.96]{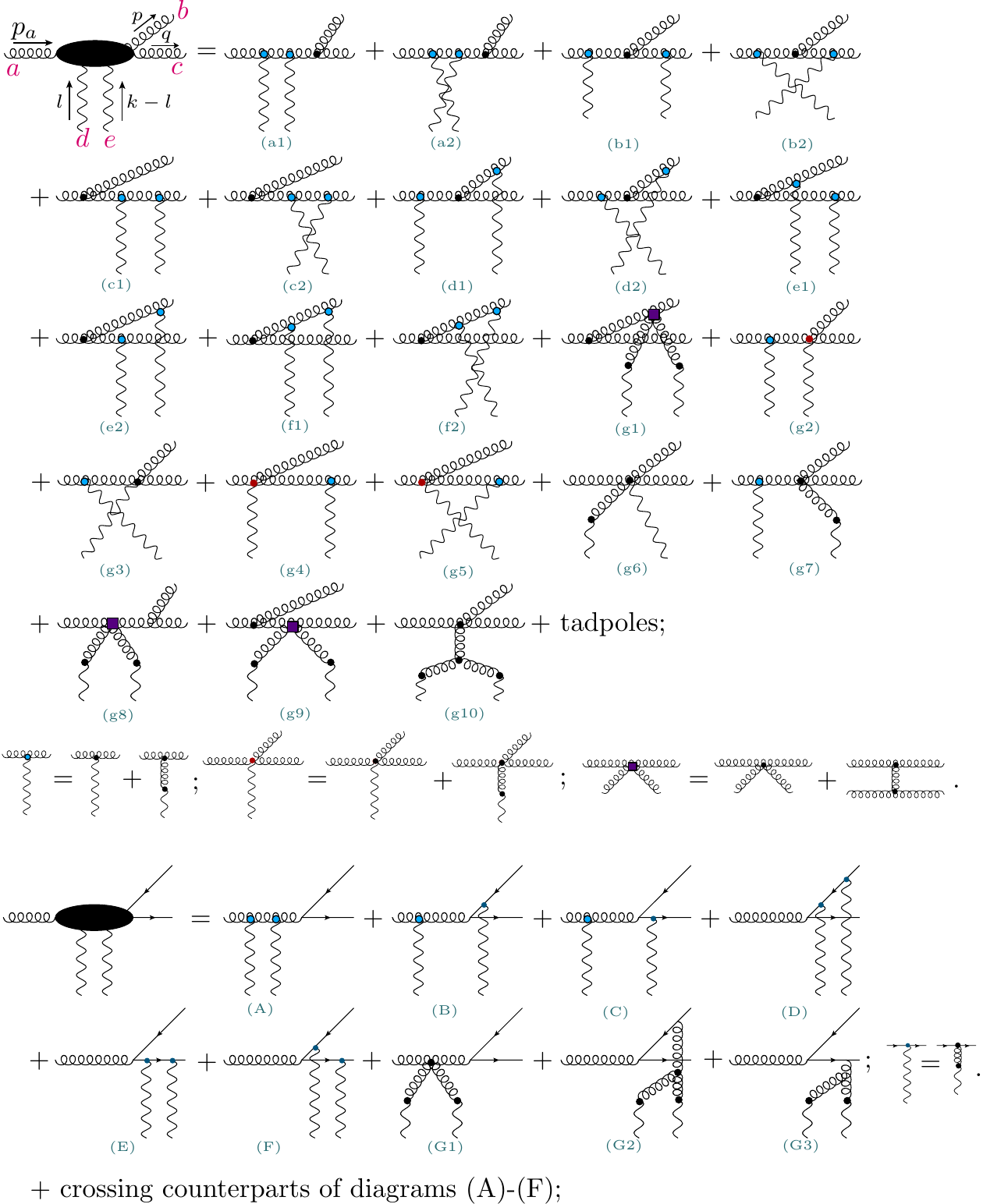}
\caption{Summary of the NLO quasielastic corrections, including $gg$ and $q\bar{q}$ final states. Tadpoles and diagrams labelled (g$i$) and (G$i$) are identically zero.}
\label{diagrams}
\end{figure}

The Feynman diagrams to be evaluated within the effective action formalism are shown in Fig. \ref{diagrams}. In analogy with Eqs.~\eqref{effenberg} and \eqref{linke}, we can write
\begin{equation}\label{klisman}
\begin{aligned}
i{\cal M}^{(0)}_{g_ag_b\to ggg}&=\frac{1}{2!}\int\frac{{\rm d}^dl}{(2\pi)^d}i\tilde{\cal M}^{abcde}_{g2r^*_+\to gg}i\tilde{\cal M}^{a'b'd'e'}_{g2r^*_-\to g}{\cal P}^{de,d'e'}\frac{(i/2)^2}{\bm{l}^2(\bm{k}-\bm{l})^2}\\&=\int\frac{{\rm d}^{2+2\epsilon}\bm{l}}{(2\pi)^{2+2\epsilon}}\frac{\phi_{ggg,{\rm a}}\,\phi_{gg,{\rm b}}}{\bm{l}^2(\bm{k}-\bm{l})^2},
\end{aligned}
\end{equation}
with a similar formula for the $q\bar{q}$ final state. For the $gg$ final state, we have
\begin{equation}
i\phi_{ggg}=\int\frac{{\rm d}l^-}{8\pi}i{\cal M}^{abcde}_{g2r^*\to gg}P^{de}=ig^3 p_a^+\frac{N_c}{\sqrt{N_c^2-1}}f_{abc}\,\varepsilon_a^\lambda\varepsilon_b^{*\mu}\varepsilon_c^{*\nu}\sum_{i=\{{\rm a,\cdots, f}\}}{\cal M}_{\lambda\mu\nu, i}.
\end{equation}
Here, $\varepsilon_a\equiv\varepsilon(p_a),\,\varepsilon_b\equiv\varepsilon(p),\,\varepsilon_c\equiv\varepsilon (q)$. The label $i$ in the subamplitudes ${\cal M}_{\lambda\mu\nu, i}$ matches the notation used in Fig.~\ref{diagrams}. The evaluation of the integral over $l^-$ is carried out using the residue theorem, taking at a time those contributions related by crossing of the external reggeons so as to ensure the vanishing of the integral over the contour at infinity. Any potentially dangerous numerators involving $l^-$, that would produce a non-zero contribution at infinity, vanish since in the numerators the momentum $l$ appears always contracted with some polarization vector, and in our gauge they satisfy $\varepsilon\cdot n^+=0$, {\it i.e.} $\varepsilon\cdot l$ does not give rise to any factor of $l^-$ (note that $l=l^-\frac{n^+}{2}+{\bm l}$).\\

After this integration, the non-vanishing subamplitudes are
\begin{equation}\label{amps}
\begin{aligned}
{\cal M}_{\lambda\mu\nu, ({\rm a})}&=-\frac{z(1-z)}{\bm{\Delta}^2}\left[\left((1-2z)k-p+q\right)_\lambda \,g_{\mu\nu}+(k+p+p_{a})_\nu \,g_{\lambda\mu}-(k+p_{a}+q)_\mu \,g_{\nu\lambda}\right],\\
{\cal M}_{\lambda\mu\nu, ({\rm c})}&=\frac{1}{2(p_a-p)^2}\left[\left((z-2)k-(z+2)p+(2-3z)p_a\right)_\nu\,g_{\lambda\mu}+4z(p_\lambda\,g_{\mu\nu}+p_{a\mu}\,g_{\nu\lambda})\right],\\
{\cal M}_{\lambda\mu\nu, ({\rm f})}&=\frac{1}{2(p_a-q)^2}\left[\left((1+z)k+(1-3z)p_a-(z-3)q\right)_\mu\,g_{\nu\lambda}-4(1-z)(p_{a\nu}\,g_{\lambda\mu}+q_\lambda\,g_{\mu\nu})\right],\\
{\cal M}_{\lambda\mu\nu, ({\rm e1})}&=\frac{1}{2\bm{\Upsilon}_i^2}\big[z(1-z)(k-2l_i+p-q)_\lambda\,g_{\mu\nu}+z\left\{(1-z)(-k+p_a+q)+2l_i\right\}_\mu\,g_{\nu\lambda}\\
&+(1-z)\left\{-z(p+p_a)+(z-2)k+2l_i\right\}_\nu\,g_{\lambda\mu}\big],\\
{\cal M}_{\lambda\mu\nu, ({\rm e2})}&=\frac{1}{2\bm{\Sigma}_i^2}\big[z(1-z)(-k+2l_i+p-q)_\lambda\,g_{\mu\nu}+z\left\{(1-z)(p_a+q)+(1+z)k-2l_i\right\}_\mu\,g_{\nu\lambda}\\&+(1-z)\left\{-z(p+p_a)+zk-2l_i\right\}_\nu\,g_{\lambda\mu}\big],
\end{aligned}
\end{equation}
where $l_i~(i=1,2)$, are the reggeon loop momenta, with $i=1$ assigned to the amplitude and $i=2$ for the complex conjugate. We have defined
\begin{equation}
\bm{\Delta}=\bm{q}-z\bm{k},\qquad \bm{\Sigma}_i=\bm{q}-\bm{l}_i,\qquad \bm{\Upsilon}_i=\bm{q}-\bm{k}+\bm{l}_i=\bm{l}_i-\bm{p},\qquad i=1,2.
\end{equation}
Using that, with our choice of polarization vectors,
\begin{equation}
\Gamma_{gr^*\to g}^{abf}=\parbox{2.4cm}{\includegraphics[width=2.5cm]{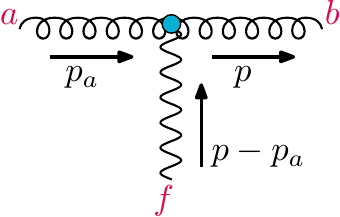}} = 2gp_a^+f_{abf}\,\varepsilon_a\cdot\varepsilon_b^*,
\end{equation}
it is possible to show that
\begin{equation}
\lim_{s_{gg}\to 0}i\phi_{ggg}= \lim_{z\to 0}\Gamma^{abf}_{gr^*\to g}a_{r^*2r^*\to g}^{fc}(\bm{p},\bm{q},\bm{l}_1)+\lim_{z\to 1}\Gamma^{acf}_{gr^*\to g}a^{fb}_{r^*2r^*\to g}(\bm{q},\bm{p},\bm{l}_1),
\end{equation}
where
\begin{equation}
\begin{aligned}
a^{fc}_{r*2r^*\to g}(\bm{p},\bm{q},\bm{l})&=\int\frac{{\rm d}l^-}{8\pi}\parbox{3.1cm}{\includegraphics[width=3cm]{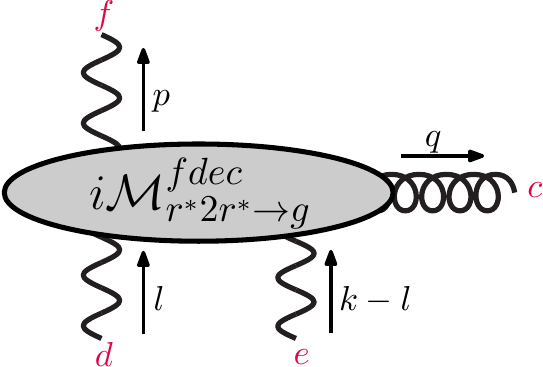}}P^{de}\\&=-\frac{g^2C_a\delta^{fc}\bm{p}^2}{\sqrt{N_c^2-1}}\bm{\varepsilon}(q)\cdot\left[2\frac{\bm{p}}{\bm{p}^2}-\frac{(\bm{p}-\bm{l})}{(\bm{p}-\bm{l})^2}+\frac{(\bm{q}-\bm{l})}{(\bm{q}-\bm{l})^2}\right],
\end{aligned}
\end{equation}
is the central production vertex (see Eq.~(93) from \cite{quark}). This indicates that the result \eqref{amps} is in agreement with rapidity factorization.\\

In the same way, we have for the diagrams with quark-antiquark final state
\begin{equation}
i\phi_{gq\bar{q}}=\int\frac{{\rm d}l^-}{8\pi}i{\cal M}^{ade}_{g2r^*\to q\bar{q}}P^{de}=\frac{g^3 t^a}{\sqrt{N_c^2-1}}\varepsilon_{a\mu}\bar{u}(p)\left[\sum_{j=\{{\rm A,\cdots,F}\}}\Gamma^\mu_j\right] v(q),
\end{equation}
with the non-vanishing subamplitudes (after the $l^-$ integration) reading in this case
\begin{equation}\label{qq}
\begin{aligned}
\Gamma^\mu_{\rm (A)}&=\frac{C_a\,z(1-z)}{\bm{\Delta}^2}[p_a^+\gamma^\mu-k^\mu\slashed{n}^+],\qquad
\Gamma^\mu_{\rm (D)}=\frac{C_f}{2(p-k)^2}[\slashed{n}^+\slashed{p}_a\gamma^\mu-2q^\mu\slashed{n}^+],\\
\Gamma^\mu_{\rm (E)}&=\frac{C_f}{2(q-k)^2}[2zp_a^+\gamma^\mu-\gamma^\mu\slashed{k}\slashed{n}^+],\\
\Gamma^\mu_{\rm (F1)}&=\frac{2C_f-C_a}{4\bm{\Upsilon}_1^2}\left\{(1-z)[\gamma^\mu(\slashed{k}-\slashed{l}_1)\slashed{n}^+-2zp_a^+\gamma^\mu]+z\slashed{l}_1\gamma^\mu\slashed{n}^+\right\},\\
\Gamma^\mu_{\rm (F2)}&=\frac{2C_f-C_a}{4\bm{\Sigma}_1^2}\left\{(1-z)[\gamma^\mu\slashed{l}_1\slashed{n}^+-2zp_a^+\gamma^\mu]+z(\slashed{k}-\slashed{l}_1)\gamma^\mu\slashed{n}^+\right\}.
\end{aligned}
\end{equation}
One can check, using {\it e.g.} spinor-helicity techniques \cite{elvang}, that both $z\to 0$ and $z\to 1$ limits of the expressions appearing in \eqref{qq} are suppressed at least as $\sqrt{z}$ (respectively $\sqrt{1-z}$), in agreement again with high-energy factorization.\\

The 3-particle phase space is
\begin{equation}
\begin{aligned}
\int {\rm d}\Pi_3&=\iiint\frac{{\rm d}^d p}{(2\pi)^{d-1}}\frac{{\rm d}^d q}{(2\pi)^{d-1}}\frac{{\rm d}^d p_2}{(2\pi)^{d-1}}\delta(p^2)\delta(q^2)\delta(p_2^2)(2\pi)^d\delta^d(p_a+p_b-p-q-p_2)\\&=\frac{1}{(2\pi)^{5+4\epsilon}}\iint {\rm d}^dk\,{\rm d}^dq\,\delta((p_a+k-q)^2)\delta(q^2)\delta((p_b-k)^2).
\end{aligned}
\end{equation}
Putting ${\rm d}^dk=\frac{1}{2}{\rm d}k^+{\rm d}k^-{\rm d}^{d-2}\bm{k}$, and using that $\delta((p_b-k)^2)=\frac{1}{p_b^-}\delta\bigg(k^+-\frac{\bm{k}^2}{p_b^-}\bigg)$; $\delta(q^2)=\frac{1}{zp_a^+}\delta\bigg(q^--\frac{\bm{q}^2}{zp_a^+}\bigg)$; and $\delta((p_a+k-q)^2)=\frac{1}{(1-z)p_a^+}\delta\bigg(k^--\left(\bm{k}^2+\frac{\bm{q}^2}{z}\right)\bigg)$, we get
\begin{equation}
\int{\rm d}\Pi_3=\frac{1}{4(2\pi)^{5+4\epsilon}p_a^+p_b^-}\iiint {\rm d}^{2+2\epsilon}\bm{k}\,{\rm d}^{2+2\epsilon}\bm{q}\frac{\rmd z}{z(1-z)}.
\end{equation}
Now, the contribution to the cross-section of the quasielastic real corrections is 
\begin{equation}
\begin{aligned}
{\rm d}\hat{\sigma}_{ab}^{\{ggg,gq\bar{q}\}}&=\frac{1}{2p_a^+p_b^-}\overline{|{\cal M}_{g_ag_b\to\{ggg,gq\bar{q}\}}^{(0)}|^2}\,{\rm d}\Pi_3=\frac{1}{2(2\pi)^{5+4\epsilon}(2p_a^+p_b^-)^2}\\&\times\iiint {\rm d}^{2+2\epsilon}\bm{k}\,{\rm d}^{2+2\epsilon}\bm{q}\frac{\rmd z}{z(1-z)}\iint\frac{{\rm d}^{2+2\epsilon}\bm{l}_1}{(2\pi)^{2+2\epsilon}}\frac{{\rm d}^{2+2\epsilon}\bm{l}_2}{(2\pi)^{2+2\epsilon}}\frac{\overline{|\phi_{\{ggg,gq\bar{q}\},{\rm a}}|^2}~\overline{|\phi_{gg,{\rm b}}|^2}}{\bm{l}_1^2(\bm{k}-\bm{l}_1)^2\bm{l}_2^2(\bm{k}-\bm{l}_2)^2}\\&\stackrel{\eqref{kreutz}}{\equiv} \iiint\frac{{\rm d}^{2+2\epsilon}\bm{l}_1}{\pi^{1+\epsilon}}\frac{{\rm d}^{2+2\epsilon}\bm{l}_2}{\pi^{1+\epsilon}}\frac{1}{\bm{l}_1^2(\bm{k}-\bm{l}_1)^2\bm{l}_2^2(\bm{k}-\bm{l}_2)^2}h^{(1)}_{r,\{gg,q\bar{q}\},{\rm a}}h^{(0)}_{gg,{\rm b}}{\rm d}^{2+2\epsilon}\bm{k},
\end{aligned}
\end{equation}
from which we get, using \eqref{krauss},
\begin{equation}
h^{(1)}_{r,\{ggg,q\bar{q}g,{\rm a}\}}=\frac{1}{2^\epsilon(2\pi)^{5+4\epsilon}8(p_a^+)^2}\iint {\rm d}^{2+2\epsilon}\bm{q}\,\frac{{\rm d}z}{z(1-z)}\overline{|\phi_{\{ggg,gq\bar{q}\},{\rm a}}|^2}.
\end{equation}
Introducing a cutoff on the partonic diffractive mass \eqref{imass},
\begin{equation}
\hat{M}_X^2=\frac{\bm{\Delta}^2}{z(1-z)}<\hat{M}_{X,{\rm max}}^2=xM_{X,{\rm max}}^2-(1-x)\bm{k}^2,
\end{equation}
which is equivalent to putting a cutoff $M_{X,{\rm max}}^2$ on the hadronic diffractive mass $M_X^2=(p_A+k)^2$, we get for $ggg$ final state
\begin{equation}\label{hg}
h_{r,ggg}^{(1)}=\frac{h_g^{(0)}}{2!}\frac{\alpha_{s,\epsilon}}{2\pi}\frac{1}{\mu^{2\epsilon}\Gamma(1-\epsilon)}\iint\frac{{\rm d}^{2+2\epsilon}\bm{q}}{\pi^{1+\epsilon}}\,{\rm d}z\,P_{gg}(z,\epsilon)\Theta\left[\hat{M}_{X,{\rm max}}^2-\tfrac{\bm{\Delta}^2}{z(1-z)}\right]J(\bm{q},\bm{k},\bm{l}_1,\bm{l}_2),
\end{equation}
where the additional factor $1/2!$ is introduced to account for the indistinguishability of final state gluons. In \eqref{hg},
\begin{equation}\label{parisi}
P_{gg}(z,\epsilon)=2C_a\frac{(1-z(1-z))^2}{z(1-z)}=2C_a\left[\frac{z}{1-z}+\frac{1-z}{z}+z(1-z)\right]
\end{equation}
is the real part of the gluon-gluon Altarelli-Parisi splitting function (in any number of dimensions) and
\begin{equation}\label{jota} 
\begin{aligned}
J(\bm{q},\bm{k},\bm{l}_1,\bm{l}_2)&=\left[\frac{\bm{\Delta}}{\bm{\Delta}^2}-\frac{\bm{q}}{\bm{q}^2}-\frac{\bm{p}}{\bm{p}^2}-\frac{1}{2}\left(\frac{\bm{\Sigma}_1}{\bm{\Sigma}_1^2}+\frac{\bm{\Upsilon}_1}{\bm{\Upsilon}_1^2}\right)\right]\cdot\bigg[\{1\leftrightarrow 2\}\bigg]\\&=\bm{k}^2\left(\frac{z^2}{\bm{\Delta}^2\bm{q}^2}+\frac{(1-z)^2}{\bm{\Delta}^2\bm{p}^2}-\frac{1}{\bm{p}^2\bm{q}^2}\right)+\frac{1}{4}\bigg\{\bm{l}_1^2\bigg(\frac{1}{\bm{p}^2\bm{\Upsilon}_1^2}+\frac{1}{\bm{q}^2\bm{\Sigma}_1^2}\bigg)\\&+(\bm{k}-\bm{l}_1)^2\bigg(\frac{1}{\bm{p}^2\bm{\Sigma}_1^2}+\frac{1}{\bm{q}^2\bm{\Upsilon}_1^2}\bigg)-\frac{(\bm{l}_1-z\bm{k})^2}{\bm{\Delta}^2\bm{\Sigma}_1^2}-\frac{(\bm{l}_1-(1-z)\bm{k})^2}{\bm{\Delta}^2\bm{\Upsilon}_1^2}\\&-\frac{1}{2}\frac{(\bm{k}-\bm{l}_1-\bm{l}_2)^2}{\bm{\Sigma}_1^2\bm{\Upsilon}_2^2}+\{1\leftrightarrow 2\}\bigg\}-\frac{1}{8}(\bm{l}_1-\bm{l}_2)^2\left(\frac{1}{\bm{\Sigma}_1^2\bm{\Sigma}_2^2}+\frac{1}{\bm{\Upsilon}_1^2\bm{\Upsilon}_2^2}\right).
\end{aligned}
\end{equation}
One can see that the structures appearing in \eqref{jota} parallel those obtained for the quark case in \cite{quark}. For the $q\bar{q}g$ final state, we get a similar result
\begin{equation}\label{jotapena}
\begin{aligned}
h_{r,q\bar{q}g}^{(1)}=\frac{h_g^{(0)}}{C_a^2}\frac{\alpha_{s,\epsilon}}{2\pi}\iint\frac{{\rm d}^{2+2\epsilon}\bm{q}}{\pi^{1+\epsilon}}\,{\rm d}z\,\frac{P_{qg}(z,\epsilon)}{\mu^{2\epsilon}\Gamma(1-\epsilon)}\Theta\bigg[\hat{M}_{X,{\rm max}}^2-\frac{\bm{\Delta}^2}{z(1-z)}\bigg]\tilde{J}(\bm{q},\bm{k},\bm{l}_1,\bm{l}_2),
\end{aligned}
\end{equation}
with
\begin{equation}
P_{qg}(z,\epsilon)=\frac{1}{2}\bigg[1-\frac{2z(1-z)}{1+\epsilon}\bigg],
\end{equation}
and
\begin{equation}\label{jotaparo}
\begin{aligned}
\tilde{J}(\bm{q},\bm{k},\bm{l}_1,\bm{l}_2)&=\bigg[C_a\frac{\bm{\Delta}}{\bm{\Delta}^2}-C_f\bigg(\frac{\bm{q}}{\bm{q}^2}+\frac{\bm{p}}{\bm{p}^2}\bigg)-\frac{2C_f-C_a}{2}\bigg(\frac{\bm{\Sigma}_1}{\bm{\Sigma}_1^2}+\frac{\bm{\Upsilon}_1}{\bm{\Upsilon}_1^2}\bigg)\bigg]\cdot\bigg[\{1\leftrightarrow 2\}\bigg]\\&=\bm{k}^2\bigg(C_aC_f\bigg[\frac{z^2}{\bm{\Delta}^2\bm{q}^2}+\frac{(1-z)^2}{\bm{\Delta}^2\bm{p}^2}\bigg]-\frac{C_f^2}{\bm{p}^2\bm{q}^2}\bigg)\\&+\frac{2C_f-C_a}{4}\bigg\{C_f\bigg[\bm{l}_1^2\bigg(\frac{1}{\bm{p}^2\bm{\Upsilon}_1^2}+\frac{1}{\bm{q}^2\bm{\Sigma}_1^2}\bigg)+(\bm{k}-\bm{l}_1)^2\bigg(\frac{1}{\bm{p}^2\bm{\Sigma}_1^2}+\frac{1}{\bm{q}^2\bm{\Upsilon}_1^2}\bigg)\bigg]\\&-C_a\bigg[\frac{(\bm{l}_1-z\bm{k})^2}{\bm{\Delta}^2\bm{\Sigma}_1^2}+\frac{(\bm{l}_1-(1-z)\bm{k})^2}{\bm{\Delta}^2\bm{\Upsilon}_1^2}\bigg]+\{1\leftrightarrow 2\}\bigg\}-\frac{(2C_f-C_a)^2}{8}\\&\times\bigg\{(\bm{l}_1-\bm{l}_2)^2\bigg(\frac{1}{\bm{\Sigma}_1^2\bm{\Sigma}_2^2}+\frac{1}{\bm{\Upsilon}_1^2\bm{\Upsilon}_2^2}\bigg)+(\bm{k}-\bm{l}_1-\bm{l}_2)^2\bigg(\frac{1}{\bm{\Sigma}_1^2\bm{\Upsilon}_2^2}+\frac{1}{\bm{\Sigma}_2^2\bm{\Upsilon}_1^2}\bigg)\bigg\}.
\end{aligned}
\end{equation}
Notice that for the $q\bar{q}g$ final state no divergence appears as $z\to 0$ or $z\to 1$, as expected, while the situation is different in the $ggg$ final state, due to the poles in the splitting function \eqref{parisi}.
\section{The Jet Vertex for Gluon-Initiated Jets with Rapidity Gap}\label{4}

\subsection{Virtual Corrections and Renormalization}\label{41}
The one-loop virtual corrections to the $gr^*r^*\to g$ amplitude have been already computed in \cite{imppart2}, where use is made of unitarity techniques. Here we rewrite the results, translating them into our normalization conventions, for the sake of completeness. We have
\begin{align}\label{tochovirtual}
h^{(1)}_v(\bm{k},\bm{l}_1,\bm{l}_2)&=\frac{h^{(0)}_g}{4\pi}\frac{\alpha_{s,\epsilon}\Gamma^2(1+\epsilon)}{(-\epsilon)\Gamma(1+2\epsilon)}\left[h_{v,a}^{(1)}(\bm{k},\bm{l}_1,\bm{l}_2)+h_{v,b}^{(1)}(\bm{k},\bm{l}_1,\bm{l}_2)+h_{v,c}^{(1)}(\bm{k},\bm{l}_1,\bm{l}_2)\right];\notag\\h_{v,a}^{(1)}(\bm{k},\bm{l}_1,\bm{l}_2)&=C_a\bigg[\ln\frac{s_0}{\bm{l}_1^2}\bigg(\frac{\bm{l}_1^2}{\mu^2}\bigg)^\epsilon+\ln\frac{s_0}{(\bm{k}-\bm{l}_1)^2}\bigg(\frac{(\bm{k}-\bm{l}_1)^2}{\mu^2}\bigg)^\epsilon\notag\\&+\bigg(\bigg(\frac{\bm{l}_1^2}{\mu^2}\bigg)^\epsilon+\bigg(\frac{(\bm{k}-\bm{l}_1)^2}{\mu^2}\bigg)^\epsilon\bigg)\bigg\{\frac{2}{\epsilon}-\frac{11+9\epsilon}{2(1+2\epsilon)(3+2\epsilon)}\notag\\&+\frac{n_f}{N_c}\frac{(1+\epsilon)(2+\epsilon)-1}{(1+\epsilon)(1+2\epsilon)(3+2\epsilon)}+\psi(1)+\psi(1-\epsilon)-2\psi(1+\epsilon)\bigg\}+\{\bm{l}_1\leftrightarrow\bm{l}_2\}\bigg];\notag\\h_{v,b}^{(1)}(\bm{k},\bm{l}_1,\bm{l}_2)&=2\frac{n_f}{N_c}\bigg[C_f\frac{2(1+\epsilon)^2+\epsilon}{(1+\epsilon)(1+2\epsilon)(3+2\epsilon)}\bigg\{\bigg(\frac{\bm{k}^2}{\mu^2}\bigg)^\epsilon-\bigg(\frac{\bm{l}_1^2}{\mu^2}\bigg)^\epsilon-\bigg(\frac{(\bm{k}-\bm{l}_1)^2}{\mu^2}\bigg)^\epsilon\bigg\}\notag\\&-\frac{1}{2N_c}\bigg\{\frac{2+\epsilon}{(1+\epsilon)(3+2\epsilon)}\bigg[\bigg(\frac{\bm{l}_1^2}{\mu^2}\bigg)^\epsilon+\bigg(\frac{(\bm{k}-\bm{l}_1)^2}{\mu^2}\bigg)^\epsilon\bigg]-K_3(\bm{l}_1)+\frac{2K_4(\bm{l}_1)}{1+2\epsilon}\bigg\}\notag\\&+\{\bm{l}_1\leftrightarrow\bm{l}_2\}\bigg];\notag\\h_{v,c}^{(1)}(\bm{k},\bm{l}_1,\bm{l}_2)&=C_a\bigg[-2\bigg\{\ln\frac{s_0}{\bm{l}_1^2}\bigg(\frac{\bm{l}_1^2}{\mu^2}\bigg)^\epsilon+\ln\frac{s_0}{(\bm{k}-\bm{l}_1)^2}\bigg(\frac{(\bm{k}-\bm{l}_1)^2}{\mu^2}\bigg)^\epsilon-\ln\frac{s_0}{\bm{k}^2}\bigg(\frac{\bm{k}^2}{\mu^2}\bigg)^\epsilon\notag\\&+\frac{3}{2\epsilon}-\frac{11+8\epsilon}{(1+2\epsilon)(3+2\epsilon)}-\psi(1+2\epsilon)-\psi(1+\epsilon)+\psi(1-\epsilon)+\psi(1)\bigg)\notag\\&\times\bigg(\bigg(\frac{\bm{l}_1^2}{\mu^2}\bigg)^\epsilon+\bigg(\frac{(\bm{k}-\bm{l}_1)^2}{\mu^2}\bigg)^\epsilon-\bigg(\frac{\bm{k}^2}{\mu^2}\bigg)^\epsilon\bigg)\bigg\}+2\epsilon K_1(\bm{l}_1)\notag\\&-\bigg(\frac{1}{\epsilon}+2\psi(1+2\epsilon)-2\psi(1+\epsilon)+2\psi(1-\epsilon)-2\psi(1)\bigg)\bigg(\frac{\bm{k}^2}{\mu^2}\bigg)^\epsilon\notag\\&-\frac{11+8\epsilon}{(1+2\epsilon)(3+2\epsilon)}\bigg(\bigg(\frac{\bm{l}_1^2}{\mu^2}\bigg)^\epsilon+\bigg(\frac{(\bm{k}-\bm{l}_1)^2}{\mu^2}\bigg)^\epsilon\bigg)\notag\\&-2K_2(\bm{l}_1)+4K_3(\bm{l}_1)-\frac{2(1+\epsilon)}{1+2\epsilon}K_4(\bm{l}_1)+\{\bm{l}_1\leftrightarrow\bm{l}_2\}\bigg],
\end{align}
where the integrals $K_i(\bm{k},\bm{l}_j),\,i=1,...,4$, are given in an expansion around $\epsilon=0$ in the formulae (A13)-(A18) of \cite{imppart2}. Notice that our expressions \eqref{tochovirtual} are considerably reduced with respect to those appearing in \cite{imppart2}, since we are already summing/averaging over final/initial helicities. In particular, the appearing structures
\begin{equation}
\frac{\bm{\varepsilon}_a\cdot\{\bm{k},\bm{l}-\bm{k}\}\,\bm{\varepsilon}_b\cdot\{\bm{k},\bm{l}-\bm{k}\}}{\{\bm{k}^2,(\bm{l}-\bm{k})^2\}}\propto\,\,\delta_{\lambda_a,\lambda_b}+\delta_{\lambda_a,-\lambda_b},
\end{equation}
vanish after taking the sum over initial and final helicities. Expanding in $\epsilon$, we have
{\small\begin{equation}\label{expansion}
\begin{aligned}
&h_v^{(1)}=\frac{h_g^{(0)}\alpha_{s,\epsilon}}{4\pi}\bigg\{-4C_a\left[\frac{1}{\epsilon^2}+\frac{1}{\epsilon}\ln\frac{\bm{k}^2}{\mu^2}\right]+\beta_0+C_a\left[\frac{8}{3}\pi^2-3-2\ln^2\frac{\bm{k}^2}{\mu^2}\right]+4\left[\beta_0+\frac{n_f}{3}\left[1+\frac{1}{C_a^2}\right]\right]\ln\frac{\bm{k}^2}{\mu^2}\\&+\bigg\{C_a\left[\ln\frac{\bm{k}^2}{\bm{l}_1^2}\ln\frac{\bm{l}_1^2}{s_0}+\ln\frac{\bm{k}^2}{(\bm{k}-\bm{l}_1)^2}\ln\frac{(\bm{k}-\bm{l}_1)^2}{s_0}+\ln^2\frac{\bm{l}_1^2}{(\bm{k}-\bm{l}_1)^2}\right]-\bigg[\beta_0+\frac{n_f}{3}\left[1+\frac{1}{C_a^2}\right]\bigg[\ln\frac{\bm{l}_1^2}{\mu^2}+\ln\frac{(\bm{k}-\bm{l}_1)^2}{\mu^2}\bigg]\bigg]\\&-\bigg[\frac{n_f}{3}\bigg[1+\frac{1}{C_a^2}\bigg]+\frac{\beta_0}{2}\bigg]\frac{(\bm{l}_1^2-(\bm{k}-\bm{l}_1)^2)}{\bm{k}^2}\ln\frac{\bm{l}_1^2}{(\bm{k}-\bm{l}_1)^2}-2\bigg[\frac{n_f}{C_a^2}+4C_a\bigg]\frac{(\bm{l}_1^2(\bm{k}-\bm{l}_1)^2)^{1/2}}{\bm{k}^2}\phi_1\sin\phi_1\\&-\frac{4}{3}\bigg[C_a+\frac{n_f}{C_a^2}\bigg]\frac{\bm{l}_1^2(\bm{k}-\bm{l}_1)^2}{(\bm{k}^2)^2}\bigg(2-\frac{[\bm{l}_1^2-(\bm{k}-\bm{l}_1)^2]}{\bm{k}^2}\ln\frac{\bm{l}_1^2}{(\bm{k}-\bm{l}_1)^2}\bigg)\sin^2\phi_1-2C_a\phi_1^2\\&+\frac{1}{3}\bigg[C_a+\frac{n_f}{C_a^2}\bigg]\frac{(\bm{l}_1^2(\bm{k}-\bm{l}_1)^2)^{1/2}}{(\bm{k}^2)^2}\bigg(4\bm{k}^2-12(\bm{l}_1^2(\bm{k}-\bm{l}_1)^2)^{1/2}\phi_1\sin\phi_1-(\bm{l}_1^2-(\bm{k}-\bm{l}_1)^2)\ln\frac{\bm{l}_1^2}{(\bm{k}-\bm{l}_1)^2}\bigg)\cos\phi_1\\&+\frac{16}{3}\bigg[C_a+\frac{n_f}{C_a^2}\bigg]\frac{(\bm{l}_1^2(\bm{k}-\bm{l}_1)^2)^{3/2}}{(\bm{k}^2)^3}\phi_1\sin^3\phi_1+\{\bm{l}_1\leftrightarrow\bm{l}_2,\phi_1\leftrightarrow\phi_2\}\bigg\}\bigg\}+{\cal O}(\epsilon).
\end{aligned}
\end{equation}}
Here
\begin{equation}
\phi_i=\arccos\frac{\bm{k}^2-\bm{l}_i^2-(\bm{k}-\bm{l}_i)^2}{2|\bm{l}_i||\bm{k}-\bm{l}_i|},\qquad i=1,2,
\end{equation}
is the angle between the reggeized gluon momenta, with $|\phi_{1,2}|\le\pi$, and $\beta_0=\frac{11}{3}C_a-\frac{2}{3}n_f$.

\subsection{Inclusion of the Jet Function and Counterterms and the LO Jet Vertex}\label{42}

As it occurs in general when evaluating higher-order QCD cross sections, we have come across different kinds of singularities, expressed in our case through poles in the dimensional regularization parameter $\epsilon$. The ultraviolet singularities present in the virtual contributions are removed by coupling renormalization, which amounts to adding to the cross-section the so-called UV counterterm\footnote{When looking at \eqref{expansion}, no pole of the form $\beta_0/\epsilon$ seems to appear. Actually, the UV counterterm \eqref{uvc} cancels the contribution proportional to $\beta_0/\epsilon$ in $h_{v,a}^{(1)}$ (Eq.~\eqref{tochovirtual}), coming from the one-loop correction to the gluon-gluon-reggeon vertex which, as remarked in \cite{imppart2}, is the only ultraviolet divergence appearing in the expansion of \eqref{tochovirtual}. However, an equal contribution with different sign, this time of infrared origin, is associated to $h_{v,c}^{(1)}$ in \eqref{tochovirtual}, and hence no $\beta_0/\epsilon$ factor appears in \eqref{expansion}.}
\begin{equation}\label{uvc}
h_{\rm UV ct.}^{(1)}=h_g^{(0)}\frac{\alpha_{s,\epsilon}}{2\pi}\frac{\beta_0}{\epsilon}.
\end{equation}
Another kind of singularities come from the soft (low-momentum) and collinear (small-angle) regions in both virtual and real corrections. In order to deal with these divergences, one must properly define a jet observable, which is infrared safe and either collinear safe or collinear factorizable, so that its value is independent of the number of soft and collinear particles in the final state \cite{seymour}. This is achieved by convoluting the partonic cross section with a distribution $S_J$ ({\em jet function}), which selects the configurations contributing to the particular choice of jet definition:
\begin{equation}
\frac{{\rm d}\hat{\sigma}_J}{{\rm d}J_1 {\rm d}J_2 {\rm d}^2\bm{k}}={\rm d}\hat{\sigma}\otimes S_{J_1}S_{J_2},
\end{equation} 
with ${\rm d}J_i={\rm d}^{2+2\epsilon}\bm{k}_{J_i}{\rm d}y_{J_i},\,i=1,2$, the jet phase space and $\bm{k}$ the transverse momentum transferred in the $t$-channel. At leading order, $\bm{k}$ is equal to the transverse momentum of the jet and the jet functions are trivial, identifying each of the final state particles with one of the jets through
\begin{equation}\label{trivium}
S^{(2)}_{J_i}(\bm{p}_i,x_i)=\delta(y_i-y_{J_i})\delta^{2+2\epsilon}(\bm{p}_i-\bm{k}_{J_i})=x_i\delta\left(x_i-\frac{|\bm{k}_{J,i}|e^{y_{J,i}}}{\sqrt{s}}\right)\delta^{2+2\epsilon}(\bm{p}_i-\bm{k}_{J_i}),\quad i=1,2.
\end{equation}
At next-to-leading order, the situation is more complex, since the two partons generating the jet can be emitted collinearly, or one of them can be soft. In this case, considering the parametrization \eqref{eq:Suda_central}, the following conditions must be imposed on the jet function in order to get a finite jet cross section \cite{seymour}
\begin{align}
  \label{eq:limitsS2}
  S_J^{(3)} ({\bm p}, {\bm q}, z x, x) & \to S_J^{(2)}({\bm p}, x) &&
  {\bm q} \to 0, \,z \to 0  \notag\\
  S_J^{(3)} ({\bm p}, {\bm q}, z x, x) & \to S_J^{(2)}({\bm k}, x) &&
 \frac{ {\bm q}}{z} \to  \frac{ {\bm p}}{1-z}  \notag\\
  S_J^{(3)} ({\bm p}, {\bm q}, z x, x) & \to S_J^{(2)}({\bm k},(1-z)
  x) &&
  {\bm q}\to  0  \notag\\
  S_J^{(3)} ({\bm p}, {\bm q}, z x, x) & \to S_J^{(2)}({\bm k}, z x)
  &&
  {\bm p}\to  0,
\end{align}
together with the symmetry of $S^{(3)}$ under simultaneously swapping $\bm{p}\leftrightarrow\bm{q}$ and $z\leftrightarrow 1-z$. Including the jet function, we can then generalize \eqref{kreutz} writing the differential partonic jet cross-section as
\begin{equation}
\begin{aligned}
\frac{{\rm d}\hat{\sigma}_{J,\,ab}}{{\rm d}J_1\,{\rm d}J_2\,{\rm d}^2\bm{k}}&=\frac{1}{\pi^2}\iiiint{\rm d}\bm{l}_1\,{\rm d}\bm{l}_1'\,{\rm d}\bm{l}_2\,{\rm d}\bm{l}_2'\frac{{\rm d}\hat{V}_a(\bm{l}_1,\bm{l}_2,\bm{k},\bm{p}_{J,1},y_1,s_0)}{{\rm d}J_1}\\&\times G(\bm{l}_1,\bm{l}_1',\bm{k},\hat{s}/s_0)G(\bm{l}_2,\bm{l}_2',\bm{k},\hat{s}/s_0)\frac{{\rm d}\hat{V}_b(\bm{l}_1',\bm{l}_2',\bm{k},\bm{p}_{J,2},y_2,s_0)}{{\rm d}J_2},\quad \hat{s}=x_1\,x_2\,s.
\end{aligned}
\end{equation}
Assuming that the reggeization scale $s_0$ is defined in such a way that it does not depend on the proton momentum fractions $x_{1,2}$ of the initial partons,\footnote{A typical choice would be $\ln \hat{s}/s_0=\Delta\eta$, with $\Delta\eta$ equal to the size of the gap $\Delta y_{\rm gap}$ or to the rapidity separation of the jets $\Delta y$.} we can write
\begin{equation}
\begin{aligned}
\frac{{\rm d}\sigma_{J,pp}}{{\rm d}J_1\,{\rm d}J_2\,{\rm d}^2\bm{k}}&=\sum_{i,j=\{q_k,\bar{q}_k,g\}}^{k=1,\cdots,n_f}\int_0^1 {\rm d}x_1\int_0^1{\rm d}x_2\,f^{(\rm gap)}_{i/p}(x_1,\mu_F)f^{(\rm gap)}_{j/p}(x_2,\mu_F)\frac{{\rm d}\hat{\sigma}_{J,ij}}{{\rm d}J_1\,{\rm d}J_2\,{\rm d}^2\bm{k}}\\&=\frac{1}{\pi^2}\iiiint{\rm d}\bm{l}_1\,{\rm d}\bm{l}_1'\,{\rm d}\bm{l}_2\,{\rm d}\bm{l}_2'\frac{{\rm d}V(\bm{l}_1,\bm{l}_2,\bm{k},\bm{p}_{J,1},y_1,s_0)}{{\rm d}J_1}\\&\times G(\bm{l}_1,\bm{l}_1',\bm{k},\hat{s}/s_0)G(\bm{l}_2,\bm{l}_2',\bm{k},\hat{s}/s_0)\frac{{\rm d}V(\bm{l}_1',\bm{l}_2',\bm{k},\bm{p}_{J,2},y_2,s_0)}{{\rm d}J_2}.
\end{aligned}
\end{equation}
The superindex $^{(\rm gap)}$ over the parton distribution functions ---which we will omit in the following--- indicates that, given the interactions with the proton remnants (Fig. \ref{fig:different_contributions}, (c)), they do not coincide with the standard parton densities. In principle they can be obtained from the usual parton densities by incorporating phenomenological gap survival probability factors, or can be extracted from observables insensitive to possible soft rescatterings, like jet-gap-jet cross-sections in double-Pomeron-exchange processes \cite{trzebinski}.\\

For gluon-induced jets at leading order (Sec. \ref{31}), the jet function \eqref{trivium} is trivial and
\begin{equation}
\begin{aligned}
\frac{{\rm d}\hat{V}_g^{(0)}}{{\rm d}J}&=C_a^2 v^{(0)}S_J^{(2)}(\bm{k},x),\qquad {\rm with}~v^{(0)}=h^{(0)}(\epsilon=0)=\frac{\alpha_s^2}{N_c^2-1},~\alpha_s=\frac{g^2}{4\pi};\\
\frac{{\rm d}V_g^{(0)}}{{\rm d}J}&=\int_0^1 {\rm d}x\,f_{g/p}(x,\mu_F^2)\,h_g^{(0)}|_{\epsilon=0}\,S_J^{(2)}(\bm{p},x)=C_a^2\,v^{(0)}\,x_J\,f_{g/p}(x_J,\mu_F^2).
\end{aligned}
\end{equation}

We should emphasize that it is not trivial that the process under consideration can be described within collinear factorization. One can check, however (Sec. \ref{43}) that actually all infrared singularities of the jet cross section can be absorbed in the definition of the parton densities following the DGLAP equations. Written alternatively, all remaining $\frac{1}{\epsilon^2}$ and $\frac{1}{\epsilon}$ poles from the real and virtual corrections cancel against the following collinear counterterms (in $\overline{ \rm MS}$ scheme)
\begin{equation}\label{cec}
\begin{aligned}
\frac{{\rm d}V_{\rm col.\,ct.}^{(1)}}{{\rm d}J}&=\int_0^1{\rm d}x\,f_{g/p}(x,\mu_F^2)\frac{\rmd \hat{V}^{(1)}_{\rm col.\,ct.}}{\rmd J},\quad \frac{\rmd \hat{V}^{(1)}_{\rm col.\,ct.}}{\rmd J}=\frac{\rmd \hat{V}^{(1)}_{{\rm col.\,ct.},\,q}}{\rmd J}+\frac{\rmd \hat{V}^{(1)}_{{\rm col.\,ct.},\,g}}{\rmd J};\\
\frac{\rmd \hat{V}^{(1)}_{{\rm col.\,ct.},\,q}}{\rmd J}&=-(2n_f)\frac{\alpha_{s,\epsilon}}{2\pi}\left(\frac{1}{\epsilon}+\ln\frac{\mu_F^2}{\mu^2}\right)\int_{z_0}^1\rmd z\,S_J^{(2)}(\bm{k},zx)\,h_q^{(0)}\,P_{qg}^{(0)}(z),\\
\frac{\rmd \hat{V}^{(1)}_{{\rm col.\,ct.},\,g}}{\rmd J}&=-\frac{\alpha_{s,\epsilon}}{2\pi}\left(\frac{1}{\epsilon}+\ln\frac{\mu_F^2}{\mu^2}\right)\int_{z_0}^1\rmd z\,S_J^{(2)}(\bm{k},zx)\,h_g^{(0)}\,P_{gg}^{(0)}(z),
\end{aligned}
\end{equation}
where\footnote{The plus distribution is defined by 
\begin{equation}
\int_\alpha^1\rmd x\,f(x)[g(x)]_+\equiv\int_\alpha^1\rmd x\,(f(x)-f(1))g(x)-f(1)\int_0^\alpha\rmd x\,g(x),
\end{equation}
when acting over a function $g(x)$ which is smooth as $x\to 1$. Even though 1 is one of the integration limits, it will be understood in the following that $\int_\alpha^1\rmd z\,f(z)\,\delta(1-z)=f(1)$, with no 1/2 factor.}
\begin{equation}
P_{qg}^{(0)}(z)=\frac{1}{2}\left[z^2+(1-z)^2\right],\quad P_{gg}^{(0)}(z)=2C_a\left[\frac{z}{[1-z]_+}+\frac{1-z}{z}+z(1-z)\right]+\frac{\beta_0}{2}\delta(1-z),
\end{equation}
are the regularized leading order splitting functions. The lower limit of integration $z_0$ is determined from the implicit factor $\delta\left((zxp_A+k)^2\right)$ in \eqref{cec}, giving the partonic diffractive mass at leading order after the rescaling $x\to zx$. Then we have
\begin{equation}
\hat{M}_{X}^2=p_a^+k^--\bm{k}^2=\frac{(1-z)\bm{k}^2}{z}<\hat{M}_{X,{\rm max}}^2\Rightarrow z_0=\frac{\bm{k}^2}{\hat{M}_{X,{\rm max}}^2+\bm{k}^2}.
\end{equation}
\subsection{Cancellation of Soft and Collinear Divergences}\label{43}
In order to explicitly check the finiteness of the jet cross section after reabsorption of the singular terms in the renormalization of the coupling and the parton distribution functions, we will need to isolate the singular regions giving rise to poles in $\epsilon$ in the phase space integrals appearing in the real emission corrections, Eqs. \eqref{hg} and \eqref{jotapena}. To this effect, we will introduce a phase slicing parameter $\lambda^2\to 0$ \cite{fabricius}; the final finite result for the jet vertex $\frac{\rmd \hat{V}^{(1)}}{\rmd J}$ will depend on $\lambda^2$ but it should be kept in mind that $\frac{\rmd}{\rmd \lambda}\frac{\rmd \hat{V}^{(1)}}{\rmd J}\to 0$ for $\lambda^2\ll \bm{k}^2$.\\

The NLO jet vertex will be the sum of several contributions
\begin{equation}
\begin{aligned}
\frac{\rmd V^{(1)}_g}{\rmd J}&=\int_0^1\rmd x\,f_{g/p}(x,\mu_F^2)\frac{\rmd\hat{V}_g^{(1)}}{\rmd J};\\\frac{\rmd\hat{V}_g^{(1)}}{\rmd J}&=\frac{\rmd\hat{V}_{v}^{(1)}}{\rmd J}+\frac{\rmd\hat{V}_{r}^{(1)}}{\rmd J}+\frac{\rmd\hat{V}_{\rm UV \,ct.}^{(1)}}{\rmd J}+\frac{\rmd\hat{V}_{\rm col.\,ct.}^{(1)}}{\rmd J},
\end{aligned}
\end{equation}
with
\begin{equation}
\begin{aligned}
\frac{\rmd\hat{V}_{v}^{(1)}}{\rmd J}&=h_v^{(1)}S^{(2)}_J(\bm{k},x);\quad \frac{\rmd\hat{V}_{\rm UV\,ct.}^{(1)}}{\rmd J}=h_{\rm UV\,ct.}^{(1)}S^{(2)}_J(\bm{k},x);\\
\frac{\rmd\hat{V}_{r}^{(1)}}{\rmd J}&=\frac{\rmd\hat{V}_{r,\,q\bar{q}g}^{(1)}}{\rmd J}+\frac{\rmd\hat{V}_{r,\,ggg}^{(1)}}{\rmd J},\qquad \frac{\rmd\hat{V}_{r,\{q\bar{q}g,\,ggg\}}^{(1)}}{\rmd J}=h^{(1)}_{r,\{q\bar{q}g,\,ggg\}}S^{(3)}_J(\bm{p},\bm{q},zx,z).
\end{aligned}
\end{equation}
In our study of the singularities of the real contribution, the following integrals will be useful
\begin{equation}\label{inte}
\int\frac{\rmd^{2+2\epsilon} \bm{q}}{\pi^{1+\epsilon}}\frac{\Theta(\lambda^2-\bm{q}^2)}{\bm{q}^2}=\frac{\lambda^{2\epsilon}}{\epsilon\Gamma(1+\epsilon)};~~~\mu^{-2\epsilon}\int\frac{\rmd^{2+2\epsilon} \bm{q}}{\pi^{1+\epsilon}}\frac{\bm{k}^2}{\bm{q}^2(\bm{q}-\bm{k})^2}=\left[\frac{\bm{k}^2}{\mu^2}\right]^\epsilon\frac{\Gamma^2(\epsilon)\Gamma(1-\epsilon)}{\Gamma(2\epsilon)},
\end{equation}
as well as the identity
\begin{equation}\label{iden}
\frac{1}{(1-z)^{1-2\epsilon}}=\frac{1}{2\epsilon}\delta(1-z)+\frac{1}{[1-z]_+}+2\epsilon\left[\frac{\ln(1-z)}{(1-z)}\right]_++{\cal O}(\epsilon^2).
\end{equation}

\subsubsection*{Extraction of Singularities: ${\bm{q\bar{q}}}$ Final State}
The poles in $\epsilon$ in the expression
\begin{equation}\label{jotaquark}
\begin{aligned}
\frac{\rmd \hat{V}_{r,\,q\bar{q}g}^{(1)}}{\rmd J}&=n_f\frac{h_g^{(0)}}{C_a^2}\frac{\alpha_{s,\epsilon}}{2\pi}\frac{1}{\mu^{2\epsilon}\Gamma(1-\epsilon)}\int_0^1\rmd z\int\frac{\rmd^{2+2\epsilon}\bm{q}}{\pi^{1+\epsilon}}P_{qg}(z,\epsilon)\\&\times\tilde{J}(\bm{q},\bm{k},\bm{l}_1,\bm{l}_2)\Theta\left(\hat{M}_{X,{\rm max}}^2-\frac{\bm{\Delta}^2}{z(1-z)}\right)S^{(3)}_J(\bm{k}-\bm{q},\bm{q},zx,x),
\end{aligned}
\end{equation}
come from the regions where the denominators in $\tilde{J}(\bm{q},\bm{k},\bm{l}_1,\bm{l}_2)$ (Eq. \eqref{jotaparo}) vanish. While $\tilde{J}(\bm{q},\bm{k},\bm{l}_1,\bm{l}_2)$ is finite as $\bm{\Sigma}_{1,2}^2\to 0$ and $\bm{\Upsilon}_{1,2}^2\to 0$, it develops singularities for $\{\bm{q}^2,\bm{p}^2,\bm{\Delta}^2\}\to 0$. For fixed $\bm{k}^2$, the regions $\bm{q}^2\to 0$ and $\bm{p}^2\to 0$ cannot overlap, but we will have to take special care of the regions where simultaneously $\bm{\Delta}^2\to 0$ and $\bm{q}^2\to 0$ or $\bm{p}^2\to 0$.

We note that \eqref{jotaquark} is symmetric under the simultaneous replacement $\bm{q}\leftrightarrow\bm{p},\,z\leftrightarrow 1-z$ (remember that $\bm{\Delta}^2=(\bm{q}-z\bm{k})^2=(\bm{p}-(1-z)\bm{k})^2$). Using this symmetry, we can rewrite \eqref{jotaparo} in the following way
\begin{align}
\label{eq:eee}
\tilde{J}&(\bm{q},\bm{k},\bm{l}_1,\bm{l}_2)
= \frac{C_a^2}{2} \left[\frac{z^2 {\bm k}^2}{{\bm \Delta}^2 {\bm q}^2} 
 +
\frac{(1-z)^2 {\bm k}^2}{{\bm \Delta}^2 {\bm p}^2}
-
\frac{ {\bm k}^2}{{\bm p}^2 {\bm q}^2}
  \right]
 + C_f^2 \frac{ {\bm k}^2}{{\bm p}^2 {\bm q}^2} \notag\\
&- \frac{1}{2} \bigg[
J_{1} ({\bm q}, {\bm k}, {\bm l}_1, z)  + J_{1} ({\bm q}, {\bm k}, {\bm l}_2, z)  + J_{1} ({\bm p}, {\bm k}, {\bm l}_1, 1-z)  + J_{1} ({\bm p}, {\bm k}, {\bm l}_2,1- z) 
 \bigg] \notag \\
& + \frac{1}{2 C_a^2} \bigg[J_2(\bm{q},\bm{k},\bm{l}_1,\bm{l}_2) + J_2(\bm{p},\bm{k},\bm{l}_1,\bm{l}_2)  - \frac{{\bm k}^2}{{\bm p}^2 {\bm q}^2}\bigg],
\end{align}
where we have introduced a notation paralleling that of \cite{quark}:
\begin{align}\label{pepe}
 J_0 ({\bm q}, {\bm k}, z) & = \frac{z^2\bm{k}^2}{\bm{\Delta}^2\bm{q}^2}, \nonumber\\ 
J_{1} ({\bm q}, {\bm k}, {\bm l}_i, z)   & = \frac{1}{4}
\bigg[
 2 \frac{{\bm k}^2}{{\bm p}^2}
\bigg(\frac{(1-z)^2}{{\bm \Delta}^2} - \frac{1}{{\bm q}^2} \bigg)
-
\frac{1}{{\bm \Sigma}_i^2}
\bigg(
\frac{({\bm l}_i - z {\bm k})^2}{{\bm \Delta}^2} - 
\frac{{\bm l}_i^2}{{\bm q}^2}
\bigg) \nonumber\\
& \qquad \qquad \qquad \qquad 
-
\frac{1}{{\bm \Upsilon}_i^2}
\bigg( 
  \frac{({\bm l}_i - (1-z) {\bm k})^2}{{\bm \Delta}^2}  
-
\frac{({\bm l}_i - {\bm k})^2}{{\bm q}^2}
\bigg)
\bigg],~i=1,2;  \\\nonumber
 J_{2} ({\bm q}, {\bm k}, {\bm l}_1, {\bm l}_2)  &=
\frac{1}{4} \bigg[
\frac{{\bm l}_1^2}{ {\bm p}^2 {\bm \Upsilon}^2_1} 
+ 
\frac{( {\bm k} - {\bm l}_1)^2}{ {\bm p}^2 {\bm \Sigma}^2_1}
 +
\frac{{\bm l}_2^2}{ {\bm p}^2 {\bm \Upsilon}^2_2} 
+ 
\frac{( {\bm k} - {\bm l}_2)^2}{ {\bm p}^2 {\bm \Sigma}^2_2}
 \\\nonumber
& 
- \frac{1}{2}
\bigg(
\frac{({\bm l}_1 - {\bm l}_2)^2}{{\bm \Sigma}_1^2 {\bm \Sigma}_2^2}
+
\frac{({\bm k} - {\bm l}_1 - {\bm l}_2)^2}{ {\bm \Upsilon}_1^2 {\bm \Sigma}_2^2   }
+
\frac{({\bm k} - {\bm l}_1 - {\bm l}_2)^2}{  {\bm \Sigma}_1^2 {\bm \Upsilon}_2^2  }
+
\frac{({\bm l}_1 - {\bm l}_2)^2}{{\bm \Upsilon}_1^2 {\bm \Upsilon}_2^2}
\bigg)
 \bigg].\nonumber
\end{align}
The  function  $J_1(\bm{q},\bm{k},\bm{l}_i,z),\,i=1,2$ has  the property that it is finite  for  ${\bm q}$  collinear to ${\bm k}$ ($\bm{p}\to 0$),  $z {\bm k}$ ($\bm{\Delta}\to 0$), ${\bm l}_i$ ($\bm{\Sigma}_i\to 0$), and ${\bm  k} - {\bm l}_i$ ($\bm{\Upsilon}_i\to 0)$. In addition, $J_1(\bm{q},\bm{k},\bm{l}_i,z=0) = 0,\,i=1,2$. The function
$J_2(\bm{q},\bm{k},\bm{l}_1,\bm{l}_2)$ is also finite for all possible
collinear poles, apart from the limit ${\bm p}^2 \to 0$ where one
finds $J_2(\bm{q},\bm{k},\bm{l}_1,\bm{l}_2) \to 1/{\bm p}^2$. Note
that with this property, the last line of Eq.~\eqref{eq:eee} is also
finite for all possible collinear poles. The only possible source for
collinear poles is therefore due to the first line of
Eq.~\eqref{eq:eee}. Making use of the symmetry of the jet vertex under
$\{ {\bm q}, z \} \leftrightarrow \{ {\bm p}, 1-z \}$ we find
\begin{equation}\label{jotaparov}
\begin{aligned}
\frac{\rmd \hat{V}_{r,\,q\bar{q}g}^{(1)}}{\rmd J}
&=h^{(0)}\frac{\alpha_{s,\epsilon}}{2\pi}\frac{n_f(1+\epsilon)}{\mu^{2\epsilon}\Gamma(1-\epsilon)}\int_0^1\rmd z\int\frac{\rmd^{2+2\epsilon}\bm{q}}{\pi^{1+\epsilon}}P_{qg}(z,\epsilon)
\bigg\{\Theta\left(\hat{M}_{X,{\rm max}}^2-\frac{{\bm \Delta}^2}{z(1-z)}\right)\\ & \times S^{(3)}_J(\bm{k}-\bm{q},\bm{q},zx,x)
\bigg[ C_a^2 \bigg( \frac{z^2 {\bm k}^2}{{\bm \Delta}^2 {\bm q}^2}  - \frac{{\bm k}^2}{({\bm p}^2 + {\bm q}^2) {\bm q}^2  }  \bigg) + 
2 C_f^2 \frac{{\bm k}^2}{({\bm p}^2 + {\bm q}^2) {\bm q}^2  } 
\notag \\
&-  \bigg[
J_{1} ({\bm q}, {\bm k}, {\bm l}_1, z)  + J_{1} ({\bm q}, {\bm k}, {\bm l}_2, z)  
 \bigg] + \frac{1}{ C_a^2} \bigg[J_2(\bm{q},\bm{k},\bm{l}_1,\bm{l}_2)  - \frac{{\bm k}^2}{{\bm p}^2( {\bm p}^2 + {\bm q}^2)}\bigg]
 \bigg].
\end{aligned}
\end{equation}
The last two lines are already finite and require no further
treatment. For the term in the second line, proportional to $C_a^2$
we note that the only singularity is due to the pole in $1/{\bm
  \Delta}^2$. Special care is however needed in the limit $z \to 0$
where the $1/{\bm q}^2$ pole appears to remain uncancelled.  Similarly, for the term proportional to $C_f^2$ we only have a $1/{\bm q}^2$ collinear singularity. We therefore find
\begin{align}
  \label{eq:decompose}
  \frac{\rmd \hat{V}_{r,\,q\bar{q}g}^{(1)}}{\rmd J} &=
\frac{\rmd \hat{V}_{r,\,q\bar{q}g}^{(1),a}}{\rmd J}+
\frac{\rmd \hat{V}_{r,\,q\bar{q}g}^{(1),d}}{\rmd J},
\end{align}
with
\begin{align}\label{jotaparova}
\frac{\rmd \hat{V}_{r,\,q\bar{q}g}^{(1),a}}{\rmd J}
&=h^{(0)}\frac{\alpha_{s,\epsilon}}{2\pi}\frac{n_f(1+\epsilon)}{\mu^{2\epsilon}\Gamma(1-\epsilon)}\int_0^1\rmd z\int\frac{\rmd^{2+2\epsilon}\bm{q}}{\pi^{1+\epsilon}}P_{qg}(z,\epsilon)
\bigg\{\Theta\left(\hat{M}_{X,{\rm max}}^2-\frac{{\bm \Delta}^2}{z(1-z)}\right) 
\notag \\ &  \times  S^{(3)}_J(\bm{k}-\bm{q},\bm{q},zx,x)
\bigg[ 
-  
J_{1} ({\bm q}, {\bm k}, {\bm l}_1, z)  - J_{1} ({\bm q}, {\bm k}, {\bm l}_2, z) 
\notag \\
& \qquad \qquad \qquad \qquad 
+ \frac{1}{ C_a^2} \bigg(J_2(\bm{q},\bm{k},\bm{l}_1,\bm{l}_2)  - \frac{{\bm k}^2}{{\bm p}^2 ( {\bm p}^2 +  {\bm q}^2)}\bigg)
 \bigg]\bigg\},
\end{align}
and the divergent terms, 
\begin{equation}
\begin{aligned}
  \label{eq:d}
\frac{\rmd \hat{V}_{r,\,q\bar{q}g}^{(1),d}}{\rmd J}
&=h^{(0)}\frac{\alpha_{s,\epsilon}}{2\pi}\frac{n_f(1+\epsilon)}{\mu^{2\epsilon}\Gamma(1-\epsilon)}\int_0^1\rmd z\int\frac{\rmd^{2+2\epsilon}\bm{q}}{\pi^{1+\epsilon}}P_{qg}(z,\epsilon)
\\ &
 \times \bigg\{ C_a^2\bigg(\Theta\left(\hat{M}_{X,{\rm max}}^2-\frac{z {\bm p}^2}{(1-z)}\right)    S^{(3)}_J(\bm{k}-z\bm{q},z\bm{q},zx,x)  \frac{z^{2 \epsilon} {\bm k}^2}{{\bm p}^2 {\bm q}^2}   \\
&
-  \Theta\left(\hat{M}_{X,{\rm max}}^2-\frac{\bm{\Delta}^2}{z(1-z)}\right)    S^{(3)}_J(\bm{k}-\bm{q},\bm{q},zx,x)  \frac{{\bm k}^2}{({\bm p}^2 + {\bm q}^2) {\bm q}^2  }  \bigg) 
 \\
&
+  \Theta\left(\hat{M}_{X,{\rm max}}^2-\frac{\bm{\Delta}^2}{z(1-z)}\right)    S^{(3)}_J(\bm{k}-\bm{q},\bm{q},zx,x) 
2 C_f^2 \frac{{\bm k}^2}{({\bm p}^2 + {\bm q}^2) {\bm q}^2  } 
\bigg \},
\end{aligned}
\end{equation}
where  we rescaled ${\bm q} \to z {\bm q}$ when necessary. We find
\begin{align}
  \label{eq:contc}
  \frac{\rmd \hat{V}_{r,\,q\bar{q}g}^{(1),d}}{\rmd J} & = \frac{\rmd \hat{V}_{r,\,q\bar{q}g}^{(1),b}}{\rmd J} + \frac{\rmd \hat{V}_{r,\,q\bar{q}g}^{(1),c}}{\rmd J}
\notag \\
\frac{\rmd \hat{V}_{r,\,q\bar{q}g}^{(1),c}}{\rmd J}  &=h^{(0)}\frac{\alpha_{s,\epsilon}}{2\pi}\frac{n_f(1+\epsilon)}{\mu^{2\epsilon}\Gamma(1-\epsilon)}\int_0^1\rmd z\int\frac{\rmd^{2+2\epsilon}\bm{q}}{\pi^{1+\epsilon}}P_{qg}(z,\epsilon) z^{2\epsilon} \notag
\\  &\bigg\{
 C_a^2
 \frac{\Theta({\lambda^2 - \bm p^2}) {\bm k}^2}{({\bm p}^2 + {\bm q}^2) {\bm p}^2}\Theta\left(\hat{M}_{X,{\rm max}}^2-\frac{z {\bm p}^2}{(1-z)}\right)    S^{(3)}_J(\bm{k}-z\bm{q},z\bm{q},zx,x) \notag \\
&
+ 2 C_f^2 \Theta\left(\hat{M}_{X,{\rm max}}^2-\frac{{\bm \Delta}^2}{z(1-z)}\right)    S^{(3)}_J(\bm{k}-\bm{q},\bm{q},zx,x) 
 \frac{{\bm k}^2 \Theta(  \lambda^2 - {\bm q}^2)}{({\bm p}^2 + {\bm q}^2) {\bm q}^2  } 
\bigg\}
\notag \\
 &=h^{(0)}_g\frac{\alpha_{s,\epsilon}}{2\pi}\frac{n_f}{\Gamma(1-\epsilon)\Gamma(1 + \epsilon)\epsilon}\int_0^1\rmd z P_{qg}(z,\epsilon) z^{2\epsilon}
 C_a^2 \left(\frac{\lambda^2}{\mu^2} \right)^{\epsilon}    S^{(2)}_J(\bm{k}, x) \notag \\
&+h^{(0)}_q\frac{\alpha_{s,\epsilon}}{2\pi}\frac{2 n_f(1+\epsilon)}{\Gamma(1-\epsilon) \Gamma(1 + \epsilon)\epsilon}\int_{z_0}^1\rmd zP_{qg}(z,\epsilon) 
\left(\frac{\lambda^2}{\mu^2} \right)^{\epsilon}       S^{(2)}_J(\bm{k}, zx) 
\notag \\
&=
h^{(0)}_g\frac{\alpha_{s,\epsilon}}{2\pi} \left(\frac{n_f}{3\epsilon} + \frac{1}{3}\ln\frac{\lambda^2}{\mu^2} - \frac{5 n_f}{9} \right)    S^{(2)}_J(\bm{k}, x) \notag \\
&+h^{(0)}_q\frac{\alpha_{s,\epsilon}}{2\pi}\int_{z_0}^1\rmd z \left[  2 n_f P_{qg}^{(0)}(z) \left( \frac{1}{\epsilon} + \ln \frac{\lambda^2}{\mu^2} \right)    + n_f \right]  S^{(2)}_J(\bm{k}, zx) +{\cal O}(\epsilon),
\end{align}
and
\begin{align}
\label{eq:1ee}
\frac{\rmd \hat{V}_{r,\,q\bar{q}g}^{(1),b}}{\rmd J}  
&=h^{(0)}\frac{\alpha_{s,\epsilon}}{2\pi}\frac{n_f(1+\epsilon)}{\mu^{2\epsilon}\Gamma(1-\epsilon)}\int_0^1\rmd z\int\frac{\rmd^{2+2\epsilon}\bm{q}}{\pi^{1+\epsilon}}P_{qg}(z,\epsilon)
\notag \\   &\times 
 \Bigg\{C_a^2\bigg\{ \frac{\Theta({\bm p^2 - \lambda^2}) {\bm k}^2}{({\bm p}^2 + {\bm q}^2) {\bm p}^2}\Theta\left(\hat{M}_{X,{\rm max}}^2-\frac{z \bm{p}^2}{(1-z)}\right)    S^{(3)}_J(\bm{k}-z\bm{q},z\bm{q},zx,x)  
\notag \\
&
+\bigg[\frac{ {\bm k}^2}{({\bm p}^2 + {\bm q}^2) {\bm q}^2}\Theta\left(\hat{M}_{X,{\rm max}}^2-\frac{z \bm{p}^2}{(1-z)}\right)    S^{(3)}_J(\bm{k}-z\bm{q},z\bm{q},zx,x) 
\notag \\
&
- \Theta\left(\hat{M}_{X,{\rm max}}^2-\frac{\bm{\Delta}^2}{z(1-z)}\right)    S^{(3)}_J(\bm{p},\bm{q},zx,x)  \frac{{\bm k}^2}{({\bm p}^2 + {\bm q}^2) {\bm q}^2  } \bigg]\bigg\}
 \\
&
+ 2 C_f^2 \Theta\left(\hat{M}_{X,{\rm max}}^2-\frac{\bm{\Delta}^2}{z(1-z)}\right)    S^{(3)}_J(\bm{p},\bm{q},zx,x) 
 \frac{{\bm k}^2 \Theta({\bm q}^2 - \lambda^2)}{({\bm p}^2 + {\bm q}^2) {\bm q}^2  } \Bigg\}
\notag \\
&+
h_g^{(0)}\frac{\alpha_{s,\epsilon}}{2\pi}\frac{n_f}{\mu^{2\epsilon}\Gamma(1-\epsilon)\Gamma(1 + \epsilon)} 
\int_{z_0}^1\rmd z P_{qg}^{(0)}(z) 2 \ln (1-z) 
    S^{(2)}_J(\bm{k},zx) +{\cal O}(\epsilon).\notag
\end{align}
Note that the squared bracket in Eq.~\eqref{eq:1ee} is finite for ${\bm q}^2 \to 0$.

\subsubsection*{Extraction of Singularities: ${\bm{gg}}$ Final State}

Now we can repeat exactly the same process for the $gg$ final state. We perform the splitting
\begin{align}
  \label{eq:split_Pgg}
  P_{gg}(z, \epsilon) & =    P_{gg}^{(1)}(z, \epsilon) +   P_{gg}^{(2)}(z, \epsilon),
 \notag \\
 P_{gg}^{(1)}(z, \epsilon) & = C_a \left[ \frac{2(1-z)}{z} + z(1-z) \right],
 \qquad 
P_{gg}^{(2)}(z, \epsilon)  = C_a \left[ \frac{2z}{1-z} + z(1-z) \right].
\end{align}
We obtain
\begin{align}\label{jotaSTART}
&\frac{\rmd \hat{V}_{r,\,ggg}^{(1)}}{\rmd J}=\frac{h_g^{(0)}}{2!}\frac{\alpha_{s,\epsilon}}{2\pi}\int_0^1\rmd z\int\frac{\rmd^{2+2\epsilon}\bm{q}}{\pi^{1+\epsilon}}
\Theta\left(\hat{M}_{X,{\rm max}}^2-\frac{{\bm \Delta}^2}{z(1-z)}\right)   S^{(3)}_J(\bm{p},\bm{q},zx,x) \notag \\
&
\times\bigg\{
\frac{P_{gg}^{(1)}(z,\epsilon)}{\mu^{2\epsilon}\Gamma(1-\epsilon)}
\bigg[ J_0 (\bm{q},\bm{k}, z)  + \sum_{i=1,2}J_1(\bm{q},\bm{k},\bm{l}_i,z)
+J_2(\bm{q},\bm{k},\bm{l}_1,\bm{l}_2) \bigg] 
\notag \\ &+ 
\frac{P_{gg}^{(2)}(z,\epsilon)}{\mu^{2\epsilon}\Gamma(1-\epsilon)}
\bigg[ J_0 (\bm{p},\bm{k}, 1-z)  + \sum_{i=1,2}J_1(\bm{p},\bm{k},\bm{l}_i, 1-z)
+J_2(\bm{p},\bm{k},\bm{l}_1,\bm{l}_2) \bigg]
\bigg\}.
\end{align}
Using transformations  $z\to 1-z$,  $\bm{q}\to (1-z)\bm{q}$, and $\bm{p}\to (1-z)\bm{p}$, we find for the terms proportional to the function $J_0$
\begin{align}\label{jotaparox}
\frac{\rmd \hat{V}_{r,\,ggg}^{(1),\,[0]}}{\rmd J}&=\frac{ \alpha_{s,\epsilon} C_a {h_g^{(0)}}}{\mu^{2\epsilon}\Gamma(1-\epsilon) 2\pi}\int_0^1\rmd z\int\frac{\rmd^{2+2\epsilon}\bm{q}}{\pi^{1+\epsilon}} \left( \frac{2 z}{(1-z)^{1-2\epsilon}} + z(1-z)^{1 + 2 \epsilon} \right)  \frac{\bm{k}^2}{\bm{q}^2(\bm{q}-\bm{k})^2} 
\notag \\
 &\times\Theta  \left(\frac{\hat{M}_{X,\,{\rm max}}^2}{1-z}-\frac{(\bm{q}-\bm{k})^2}{z}\right) S^{(3)}_J(\bm{k}-(1-z)\bm{q},(1-z)\bm{q},(1-z)x,x).
\end{align}
With \eqref{iden} we have
\begin{align}
  \label{eq:idappl}
  \frac{2 z}{(1-z)^{1-2\epsilon}} + z(1-z)^{1 + 2 \epsilon} & =  \frac{1}{\epsilon} \delta(1-z) + \left\{ \frac{2z}{[(1-z)]_+} + z(1-z) \right\} 
\notag \\ &+
2 \epsilon z\left\{ (1-z) \ln(1-z) + 2 \left[ \frac{\ln(1-z)}{1-z} \right]_+ \right\}+{\cal O}(\epsilon^2).
\end{align}
For the first term we find
\begin{align}
  \label{eq:firstterm}
  \frac{\rmd \hat{V}_{r,\,ggg}^{(1), [0{\rm a}]}}{\rmd J}&={h_g^{(0)}}\frac{\alpha_{s,\epsilon} C_a}{2\pi} S_J^{(2)}({\bm k}, x) \left[\frac{2}{\epsilon^2} + \frac{2}{\epsilon} \ln \frac{{\bm k}^2}{\mu^2} +   \ln^2 \frac{{\bm k}^2}{\mu^2} - \frac{\pi^2}{3} \right]+{\cal O}(\epsilon).
\end{align}
For the  second term of \eqref{jotaparox} we use a  phase space slicing parameter to isolate the singular contributions. Separating singularities in ${
\bm q}^2$ and ${\bm p}^2$ by making use of the identity
\begin{align}
  \label{eq:fract_decmp}
  \frac{{\bm k}^2}{{\bm q}^2 {\bm p}^2} & =   \frac{{\bm k}^2}{{\bm q}^2({\bm q}^2+ {\bm p}^2)}  +   \frac{{\bm k}^2}{ {\bm p}^2 (  {\bm p}^2 + {\bm q}^2)},
\end{align}
we find
\begin{equation}
\begin{aligned}
  \label{eq:2ndtterm}
 \frac{\rmd \hat{V}_{r,\,ggg}^{(1),\,[0{\rm b}]}}{\rmd J}&  =\frac{ \alpha_{s,\epsilon} C_a {h_g^{(0)}}}{\Gamma(1-\epsilon) 2\pi} \bigg\{ \frac{(\lambda^2/\mu^2)^\epsilon}{\epsilon\Gamma(1+\epsilon)}  \bigg [ \int_{z_0}^1 \rmd z\,  S_J^{(2)} ({\bm k}, zx) \left(\frac{2z}{(1-z)_+} + z(1-z) \right)\\ & 
- \frac{11}{6} S_J^{(2)} ({\bm k}, x) 
\bigg] 
+  \int_0^1\rmd z\int\frac{\rmd^{2}\bm{q}}{\pi} 
\Theta  \left(\hat{M}_{X,\,{\rm max}}^2-\frac{(\bm{p}- z \bm{k})^2}{z (1-z)}\right)  
\\ &
 \times \left( \frac{2 z}{(1-z)_+} + z(1-z) \right) S^{(3)}_J({\bm p}, \bm{q},(1-z)x,x) 
 \Theta\left(\frac{|{\bm q}|}{1-z} - \lambda \right)
 \\&
\times\frac{ (1-z)^2{\bm k}^2}{{\bm q}^2({\bm q}^2+ (1-z)^2( {\bm p} - z {\bm k})^2)}  
+
\Theta\left( \frac{|{\bm p} - z {\bm k}|}{1-z} - \lambda \right) \\
&
   \times\frac{(1-z)^2{\bm k}^2}{( {\bm p} - z \bm{k})^2 ( (1-z)^2 ( {\bm p} - z {\bm k})^2 + {\bm q}^2)}  
 +{\cal O}(\epsilon)
\bigg\}.
\end{aligned}
\end{equation}
Finally, the third term is finite and reads
\begin{equation}
\begin{aligned}
  \label{eq:3rdterm}
\frac{\rmd \hat{V}_{r,\,ggg}^{(1),\,[0{\rm c}]}}{\rmd J} = \frac{\alpha_s}{2\pi}C_a^2 v^{(0)}&\bigg\{2C_a\int_{z_0}^1\rmd z\,S_J^{(2)}(\bm{k},zx)z\left[(1-z)\ln(1-z)+2\left[\tfrac{\ln(1-z)}{1-z}\right]_+\right]\\&+\frac{67}{18}C_a\,S_J^{(2)}(\bm{k},x)\bigg\}.
\end{aligned}
\end{equation}
The terms proportional to $J_1$ are immediately finite and require no further treatment. 
 The term with $J_2(\bm{q},\bm{k},\bm{l}_1,\bm{l}_2)$ ($J_2(\bm{p},\bm{k},\bm{l}_1,\bm{l}_2)$)  is only singular as $\bm{p}^2\to 0$ ($\bm{q}^2\to 0$). For  both scenarios the singularity at $z \to  0$ (in $P_{gg}^{(1)}(z, \epsilon)$) and  $z \to  1$ (in $P_{gg}^{(2)}(z, \epsilon)$) is regulated through the constraint on the diffractive mass.  We therefore find
\begin{align}\label{jota2}
\frac{\rmd \hat{V}_{r,\,ggg}^{(1),\,[2]}}{\rmd J}&={h_g^{(0)}}\frac{\alpha_{s,\epsilon}}{2\pi}\int_0^1\rmd z\int\frac{\rmd^{2+2\epsilon}\bm{q}}{\pi^{1+\epsilon}}
\Theta\left(\hat{M}_{X,{\rm max}}^2-\frac{{\bm \Delta}^2}{z(1-z)}\right)   S^{(3)}_J(\bm{p},\bm{q},zx,x) \notag \\
&
\hspace{5cm}
\times\frac{P_{gg}^{(1)}(z,\epsilon)}{\mu^{2\epsilon}\Gamma(1-\epsilon)}
J_2(\bm{q},\bm{k},\bm{l}_1,\bm{l}_2 )
\notag \\
&={h_g^{(0)}}\frac{\alpha_{s,\epsilon}}{2\pi} \bigg[ \int_{z_0}^1\rmd z  \frac{(\lambda^2/\mu^2)^\epsilon}{\epsilon\Gamma(1+\epsilon)}  \frac{P_{gg}^{(1)}(z,\epsilon)}{\Gamma(1-\epsilon)}S_J^{(2)}(\bm{k},zx) + 
 \int_0^1\rmd z\int\frac{\rmd^{2+2\epsilon}\bm{p}}{\pi^{1+\epsilon}} \Theta({\bm p}^2 - \lambda^2)
\notag \\
&
\times\Theta\left(\hat{M}_{X,{\rm max}}^2-\frac{{\bm \Delta}^2}{z(1-z)}\right) S^{(3)}_J(\bm{p},\bm{q},zx,x) 
\frac{P_{gg}^{(1)}(z,\epsilon)}{\mu^{2\epsilon}\Gamma(1-\epsilon)}
J_2(\bm{q},\bm{k},\bm{l}_1,\bm{l}_2 ) \bigg].
\end{align}
It is easy to see now that all poles in $\epsilon$ vanish in the final result for $\frac{\rmd \hat{V}_g^{(1)}}{\rmd J}$. The poles from the virtual terms \eqref{expansion} are cancelled by those in Eq.~\eqref{eq:firstterm}. Using that $\int_0^1\rmd z\,P_{qg}(z)=\frac{1}{3}$, and the fact that $P_{gg}^{(1)}(z,\epsilon)+C_a\left(\frac{2z}{[1-z]_+}+z(1-z)\right)=P_{gg}^{(0)}(z)-\frac{\beta_0}{2}\delta(1-z)$, one can then check that the $\frac{1}{\epsilon}$ terms proportional to $S_J^{(2)}(\bm{k},x)$ in expressions \eqref{eq:contc}, \eqref{eq:2ndtterm} and \eqref{jota2} cancel the singularity in the UV counterterm \eqref{uvc}. Similarly, the pole terms involving $S_J^{(2)}(\bm{k},zx)$ in Eqs. \eqref{eq:contc}, \eqref{eq:1ee}, \eqref{eq:2ndtterm} and \eqref{jota2} cancel against those of the collinear counterterm \eqref{cec}.
\subsection{NLO Jet Impact Factor: Final Result}\label{44}

Having checked explicitly the cancellation of singularities, we can expand and add up the former expressions to obtain the final result for the gluon-initiated jet vertex:
{\small\begin{align}\label{apocalipsis}
&\frac{\rmd\hat{V}^{(1)}(x,\bm{k},\bm{l}_1,\bm{l}_2;x_J,\bm{k}_J;M_{X,\,{\rm max}},s_0)}{\rmd J}  =v^{(0)}\frac{\alpha_s}{2\pi}\big( G_1 + G_2 + G_3 \Big);
\notag \\
& G_1  =
C_a^2\,S_J^{(2)}(\bm{k},x)
\Bigg[C_a \left( \pi^2- \frac{5}{6} \right)- \beta_0 \left( \ln\frac{\lambda^2}{\mu^2} -  \frac{4}{3} \right) 
\notag \\
&
+\left(\frac{\beta_0}{4} + \frac{11 C_a}{12}+\frac{n_f}{6 C_a^2}\right)
\left( \ln\frac{\bm{k}^4}{ {\bm l}_1^2  ({\bm k} - {\bm l}_1^2)}
+
 \ln\frac{\bm{k}^4}{ {\bm l}_2^2  ({\bm k} - {\bm l}_2)^2 } 
\right)
\notag \\
&
+\frac{1}{2}\bigg\{C_a\bigg(
\ln^2\frac{\bm{l}_1^2}{(\bm{k}-\bm{l}_1)^2}
+
\ln\frac{\bm{k}^2}{\bm{l}_1^2}\ln\frac{\bm{l}_1^2}{s_0}
+
\ln\frac{\bm{k}^2}{(\bm{k}-\bm{l}_1)^2}\ln\frac{(\bm{k}-\bm{l}_1)^2}{s_0}
\bigg)
\notag\\
&-\bigg(\frac{n_f}{3 C_a^2} + \frac{11 C_a}{6}\bigg)
\frac{\bm{l}_1^2-(\bm{k}-\bm{l}_1)^2}{\bm{k}^2}
\ln\frac{\bm{l}_1^2}{(\bm{k}-\bm{l}_1)^2}
-2\bigg(\frac{n_f}{C_a^2}+4C_a\bigg)
\notag\\
&  \times\frac{(\bm{l}_1^2(\bm{k}-\bm{l}_1)^2)^{\frac{1}{2}}}{\bm{k}^2} \phi_1\sin\phi_1 
+\frac{1}{3}\bigg(
C_a+\frac{n_f}{C_a^2}
\bigg)\bigg[
16\frac{(\bm{l}_1^2(\bm{k}-\bm{l}_1)^2)^{\frac{3}{2}}}{(\bm{k}^2)^3}\phi_1\sin^3\phi_1
\notag\\
& -4\frac{\bm{l}_1^2(\bm{k}-\bm{l}_1)^2}{(\bm{k}^2)^2}  \bigg(
2-\frac{\bm{l}_1^2-(\bm{k}-\bm{l}_1)^2}{\bm{k}^2}\ln\frac{\bm{l}_1^2}{(\bm{k}-\bm{l}_1)^2}
\bigg)
\sin^2\phi_1+\frac{(\bm{l}_1^2(\bm{k}-\bm{l}_1)^2)^{\frac{1}{2}}}{(\bm{k}^2)^2}
\notag\\ & \times\cos\phi_1
\bigg(4\bm{k}^2 -12(\bm{l}_1^2(\bm{k}-\bm{l}_1)^2)^{\frac{1}{2}}\phi_1\sin\phi_1-(\bm{l}_1^2-(\bm{k}-\bm{l}_1)^2)\ln\frac{\bm{l}_1^2}{(\bm{k}-\bm{l}_1)^2}\bigg)
\bigg]
\notag\\&-2C_a\phi_1^2+\{\bm{l}_1\leftrightarrow\bm{l}_2,\phi_1\leftrightarrow\phi_2\}\bigg\}\Bigg];
\notag \\
 & G_2 =   \int_{z_0}^1\rmd z\,S_J^{(2)}(\bm{k},zx) \bigg\{  2n_f P_{qg}^{(0)}(z)
\left(C_f^2 \ln \frac{\lambda^2}{\mu_F^2}  +  C_a^2 \ln(1-z) \right);
\notag  
 \\ &
 + C_a^2 P_{gg}^{(0)}(z) \ln\frac{\lambda^2}{\mu_F^2}  + C_f^2 n_f + 2 C_a^3 z \bigg(  (1-z)\ln(1-z)   + 2 \left[\frac{\ln (1-z)}{1-z} \right]_+ \bigg)
\notag \\
&G_3 = 
\int_0^1 \rmd z\int\frac{\rmd^2\bm{q}}{\pi} \bigg\{
n_f P^{(0)}_{qg}(z) 
\bigg[
C_a^2 \Theta\left(\hat{M}_{X,{\rm max}}^2-\frac{z {\bm p}^2}{(1-z)}\right)
\notag \\
&\times
    S^{(3)}_J(\bm{k}-z\bm{q},z\bm{q},zx,x) \bigg[ \frac{\Theta({\bm p^2 - \lambda^2}) {\bm k}^2}{({\bm p}^2 + {\bm q}^2) {\bm p}^2} +
\frac{ {\bm k}^2}{({\bm p}^2 + {\bm q}^2) {\bm q}^2} \bigg]
\notag \\
&
- \Theta\left(\hat{M}_{X,{\rm max}}^2-\frac{{\bm \Delta}^2}{z(1-z)}\right) 
  S^{(3)}_J(\bm{p},\bm{q}, zx, x)
\bigg(C_a^2  \frac{ {\bm k}^2}{({\bm p}^2 + {\bm q}^2) {\bm q}^2} 
\notag \\
&
-   
 2 C_f^2 
 \frac{{\bm k}^2 \Theta({\bm q}^2 - \lambda^2)}{({\bm p}^2 + {\bm q}^2) {\bm q}^2  } \bigg)\bigg]
+ P_1(z)  \Theta\left(\hat{ M}^2_{X,{\rm max}}-\frac{({ \bm p } - z {\bm k})^2}{z(1-z)}\right) 
\notag \\ & \times
 S_J^{(3)}({\bm p}, {\bm q}, (1-z)x, x) \frac{(1-z)^2 {\bm k}^2}{(1-z)^2 ({\bm p} - z {\bm k})^2 + {\bm q}^2}
\bigg[  \Theta\left(\frac{|\bm{q}|}{1-z}-\lambda\right)  \frac{1}{{\bm q}^2} 
\notag \\
 &+ 
\Theta\left(\frac{|\bm{p} - z {\bm k}|}{1-z}-\lambda\right)  \frac{1}{({\bm p} - z {\bm k})^2}
 + 
\Theta\left(\hat{M}_{X, \rm max}^2 - \frac{{\bm \Delta}^2}{z(1-z)} \right)
 S_J^{(3)}(\bm{p},\bm{q},zx,x) 
& \notag \\
 &
\times\bigg[
\frac{n_f }{ C_a^2}  P_{qg}^{(0)} \bigg(J_2(\bm{q},\bm{k},\bm{l}_1,\bm{l}_2)   - \frac{{\bm k}^2}{{\bm p}^2( {\bm q}^2 + {\bm p}^2)}\bigg)
- n_f  P_{qg}^{(0)} \bigg(
J_{1} ({\bm q}, {\bm k}, {\bm l}_1, z)  \notag \\
&  + J_{1} ({\bm q}, {\bm k}, {\bm l}_2, z)  
 \bigg) 
+   P_0(z) 
 \bigg( J_1(\bm{q},\bm{k},\bm{l}_1)+J_1(\bm{q},\bm{k},\bm{l}_2)+ J_2(\bm{q},\bm{k},\bm{l}_1,\bm{l}_2) \Theta(\bm{p}^2-\lambda^2)
    \bigg)\bigg] 
\bigg\}.
\end{align}}
Here, $P_{0}(z) =C_a\big[\tfrac{2(1-z)}{z}+z(1-z)\big]$, and $P_{1}(z)=C_a\big[\tfrac{2z}{[1-z]_+}+z(1-z)\big]$.
 The rest of necessary definitions appearing in \eqref{apocalipsis} are scattered throughout the text.

\section{Conclusions and Outlook}\label{5}

In this paper we have completed the analytical calculation started in \cite{quark} of the next-to-leading order corrections to the effective vertex for jet production in association to a rapidity gap, by computing the real quasielastic corrections to gluon-initiated jets. The main result is summarized in Eq.~\eqref{apocalipsis}, where the jet vertex appears as a function of a phase slicing parameter $\lambda^2$, used in the extraction of singularities, and a generic jet definition. It is interesting and nontrivial that, for the kinematics of this process that lies in the interface of collinear and BFKL-like $\bm{k}_t$ factorization, it is possible to absorb all soft and collinear singularities in the DGLAP renormalization of parton densities.\\

The result \eqref{apocalipsis} is well suited for numerical implementation using a particular jet definition. A convenient choice may be to use a cone with small radius (in the pseudorapidity-azimuthal angle plane) approximation \cite{sca}, which in the Mueller-Navelet case \cite{scmueller} provided a simple analytic result for the jet vertices projected onto the BFKL eigenfunction (($\nu, n$) representation). In addition to the jet algorithm, experimental cuts matching the future measurements at LHC \cite{lhc}, and a model of the energy dependence of the rapidity gap survival probability, must be included. 
\section*{Acknowledgments}

We thank the participants of the {\em Second Informal Meeting on Scattering Amplitudes \& the Multi-Regge Limit}, held in Madrid in February 2014, for stimulating discussions. We acknowledge partial support  by the Research Executive Agency (REA) of the European Union under the 
Grant Agreement number PITN-GA-2010-264564 (LHCPhenoNet), 
the Comunidad de Madrid through Proyecto HEPHACOS ESP-1473, 
by MICINN (FPA2010-17747),
by the Spanish Government and EU ERDF funds 
(grants FPA2007-60323, FPA2011-23778 and CSD2007- 00042 Consolider Project CPAN) 
and by GV (PROMETEUII/2013/007).
 M.H.
acknowledges support from the U.S. Department of Energy under contract
number DE-AC02-98CH10886 and a ``BNL Laboratory Directed Research and
Development'' grant (LDRD 12-034).  The research of J.D.M. is supported by the European Research Council under the
Advanced Investigator Grant ERC-AD-267258.

\appendix

\section{The Inclusive Pomeron-Gluon Impact Factor}
In this appendix, we compute the inclusive Mueller-Tang gluon-initiated jet impact factor, in the limit where the cutoff in the diffractive mass $\hat{M}_{X,\,{\rm max}}^2\to \infty$. That is, we take the jet function to be unity, and we will omit the cutoff except for those cases where the $z$-integration is not already regulated by the computation of the momentum integral in dimensional regularization. In such cases, the cutoff will be needed: keeping the cutoff finite amounts to subtracting the central production contribution. In this way we have a simple analytic check of the cancellation of singularities for the exclusive case in Sec.~\ref{4}.\\

The collinear counterterm reads in this case
\begin{equation}\label{polchinski}
\begin{aligned}
\lim_{\hat{M}_{X,\,{\rm max}}^2\to\infty}&\bigg[-\frac{\alpha_{s,\epsilon}}{2\pi}\left(\frac{1}{\epsilon}+\ln\frac{\mu_F^2}{\mu^2}\right)h^{(0)}\int_{x_0}^1\rmd x\,f_g(x,\mu_F^2)\\&\times\bigg\{C_f^2\frac{2n_f}{3}-C_a^2(1+\epsilon)\left[C_a\left(\frac{11}{3}-2\ln\frac{x}{x_0}\right)-\frac{\beta_0}{2}\right]\bigg\}\bigg],\qquad\frac{x}{x_0}=\frac{\hat{M}_{X,\,{\rm max}}^2}{\bm{k}^2}.
\end{aligned}
\end{equation}
We start evaluating the inclusive impact factor for $q\bar{q}$ final state. Symmetry allows us to substitute $\tilde{J}(\bm{q},\bm{k},\bm{l}_1,\bm{l}_2)$ by $2[\tilde{J}_0(\bm{q},\bm{k},\bm{l}_1,\bm{l}_2,z)+\tilde{J}_1(\bm{q},\bm{k},\bm{l}_1,\bm{l}_2)]$ in \eqref{jotapena}, and then use \eqref{inte} to get
\begin{equation}
\begin{aligned}
&\int\frac{\rmd^{2+2\epsilon}\bm{q}}{\pi^{1+\epsilon}}\tilde{J}(\bm{q},\bm{k},\bm{l}_1,\bm{l}_2)=2\frac{\Gamma^2(\epsilon)\Gamma(1-\epsilon)}{\Gamma(2\epsilon)}\\&\times\Bigg[C_a\bigg[C_f(z^2\bm{k}^2)^\epsilon-\frac{2C_f-C_a}{4}\{[(\bm{l}_1-z\bm{k})^2]^\epsilon+[(\bm{l}_2-z\bm{k})^2]^\epsilon\}\bigg]-\frac{C_f^2}{2}(\bm{k}^2)^\epsilon\\&\qquad +\frac{(2C_f-C_a)C_f}{4}\left[(\bm{l}_1^2)^\epsilon+(\bm{l}_2^2)^\epsilon+((\bm{k}-\bm{l}_1)^2)^\epsilon+((\bm{k}-\bm{l}_2)^2)^\epsilon\right]\\&\qquad-\frac{(2C_f-C_a)^2}{8}\left[[(\bm{l}_1-\bm{l}_2)^2]^\epsilon+[(\bm{k}-\bm{l}_1-\bm{l}_2)^2]^\epsilon\right]\Bigg].
\end{aligned}
\end{equation}
Now, we can evaluate the integral over $z$ in \eqref{jotapena} using the results\footnote{ The integrals $K_i(\bm{a},\bm{b}),\,i=1,\cdots, 4$, appearing in this appendix have been evaluated in \cite{imppart2}.}
\begin{equation}
\begin{aligned}
&\int_0^1\rmd z\,P_{qg}(z,\epsilon)=\frac{1}{3}+\frac{\epsilon}{6}+{\cal O}(\epsilon^2);\qquad\int_0^1\rmd z\,P_{qg}(z,\epsilon)(z^2)^\epsilon=\frac{1}{3}-\frac{5}{9}\epsilon+{\cal O}(\epsilon^2);\\&\int_0^1\rmd z\,P_{qg}(z,\epsilon)[(\bm{a}-z\bm{b})^2]^\epsilon=\frac{1}{2}K_3(\bm{a},\bm{b})-\frac{1}{1+\epsilon}K_4(\bm{a},\bm{b});\\&\mu^{-2\epsilon}K_3(\bm{a},\bm{b})=1+\epsilon\bigg[\frac{1}{2}\left(\ln\frac{\bm{a}^2}{\mu^2}+\ln\frac{(\bm{a}-\bm{b})^2}{\mu^2}\right)-2\\&\,\,\,\,\quad\qquad+\frac{1}{\bm{b}^2}\left[((\bm{a}-\bm{b})^2-\bm{a}^2)\ln\frac{(\bm{a}-\bm{b})^2}{\bm{a}^2}+2(\bm{a}^2(\bm{a}-\bm{b})^2)^{1/2}\vartheta\sin\vartheta\right]\bigg]+{\cal O}(\epsilon^2);\\&\mu^{-2\epsilon}K_4(\bm{a},\bm{b})=\frac{1}{6}+\epsilon\bigg[\frac{1}{12}\left(\ln\frac{\bm{a}^2}{\mu^2}+\ln\frac{(\bm{a}-\bm{b})^2}{\mu^2}\right)-\frac{5}{18}\\&\,\,\,\,\quad\qquad-\frac{2(\bm{a}^2(\bm{a}-\bm{b})^2)^{1/2}}{3(\bm{b}^2)^2} [2(\bm{a}^2(\bm{a}-\bm{b})^2)^{1/2}\sin^2\vartheta-\bm{b}^2\cos\vartheta]\\&\,\,\,\,\quad\qquad +\tfrac{(\bm{a}-\bm{b})^2-\bm{a}^2}{12(\bm{b}^2)^3}\ln\tfrac{(\bm{a}-\bm{b})^2}{\bm{a}^2}[8\bm{a}^2(\bm{a}-\bm{b})^2\sin^2\vartheta-2(\bm{a}^2(\bm{a}-\bm{b})^2)^{1/2}\bm{b}^2\cos\vartheta+(\bm{b}^2)^2]\\&\,\,\,\,\quad\qquad +\tfrac{2\bm{a}^2(\bm{a}-\bm{b})^2}{3(\bm{b}^2)^3}[4(\bm{a}^2(\bm{a}-\bm{b})^2)^{1/2}\sin^2\vartheta-3\bm{b}^2\cos\vartheta]\vartheta\sin\vartheta+{\cal O}(\epsilon^2).
\end{aligned}
\end{equation}

Here $\vartheta$ is the angle between $\bm{a}$ and $(\bm{a}-\bm{b})$, with $|\vartheta|\le\pi$. In terms of them, we have
\begin{equation}
\begin{aligned}
h^{(1)}_{r,q\bar{q}g}&=\frac{\alpha_{s,\epsilon}}{2\pi}h^{(0)}\frac{(2n_f)(1+\epsilon)\Gamma^2(\epsilon)}{\Gamma(2\epsilon)\mu^{2\epsilon}}\bigg[C_a\bigg[C_f(\bm{k}^2)^\epsilon\left(\frac{1}{3}-\frac{5\epsilon}{9}\right)\\&-\frac{2C_f-C_a}{4}\bigg\{\frac{1}{2}[K_3(\bm{l}_1,\bm{k})+K_3(\bm{l}_2,\bm{k})]-\frac{1}{1+\epsilon}[K_4(\bm{l}_1,\bm{k})+K_4(\bm{l}_2,\bm{k})]\bigg\}\bigg]\\&+\left(\frac{1}{3}+\frac{\epsilon}{6}\right)\bigg\{\frac{(2C_f-C_a)C_f}{4}\left[(\bm{l}_1^2)^\epsilon+(\bm{l}_2^2)^\epsilon+((\bm{k}-\bm{l}_1)^2)^\epsilon+((\bm{k}-\bm{l}_2)^2)^\epsilon\right]\\&-\frac{C_f^2}{2}(\bm{k}^2)^\epsilon-\frac{(2C_f-C_a)^2}{8}\left[[(\bm{l}_1-\bm{l}_2)^2]^\epsilon+[(\bm{k}-\bm{l}_1-\bm{l}_2)^2]^\epsilon\right]\bigg\}\bigg]+{\cal O}(\epsilon).
\end{aligned}
\end{equation}
Now we address the $gg$ final state (Eq.~\eqref{hg}). Using the decomposition \eqref{pepe}, we will only have to take into account the diffractive mass cutoff for the term proportional to $J_2(\bm{q},\bm{k},\bm{l}_1,\bm{l}_2)$, where the $z\to 0$ divergence is not already regulated by the momentum integral. The symmetry of $h_{r,\,ggg}^{(1)}$ under the simultaneous replacement $\{\bm{q}\leftrightarrow \bm{p},\,z\leftrightarrow 1-z\}$, allows us to substitute $P_{gg}(z,\epsilon)$ by $2P_{gg}^{(1)}(z,\epsilon)$. This substitution can be undone for the terms involving $J_0(\bm{q},\bm{k},z)$ and $J_1(\bm{q},\bm{k},\bm{l}_i,z),\,i=1,2$, since for these terms the cutoff is not needed and then they enjoy symmetry under the replacement $z\leftrightarrow 1-z$ alone. Therefore we can write, for $M_{X,\,{\rm max}}^2\to\infty$
\begin{equation}
\begin{aligned}
h_{r,\,ggg}^{(1)}&=\tfrac{h_g^{(0)}}{2!}\tfrac{\alpha_{s,\epsilon}}{2\pi}\tfrac{1}{\mu^{2\epsilon}\Gamma(1-\epsilon)}\iint\frac{\rmd^{2+2\epsilon}\bm{q}}{\pi^{1+\epsilon}}\rmd z\bigg\{P_{gg}(z,\epsilon)\bigg[J_0(\bm{q},\bm{k},z)+\sum_{i=1,2}J_1(\bm{q},\bm{k},\bm{l}_i,z)\bigg]\\&\hspace{4cm}+2P_{gg}^{(1)}(z,\epsilon)\Theta\bigg[\hat{M}_{X\,{\rm max}}^2-\tfrac{\bm{\Delta}^2}{z(1-z)}\bigg]J_2(\bm{q},\bm{k},\bm{l}_1,\bm{l}_2)\bigg\}.
\end{aligned}
\end{equation}

Using again \eqref{inte}, we have
\begin{equation}
\begin{aligned}
\textfrak{s:}_1&\equiv\int\tfrac{\rmd^{2+2\epsilon}\bm{q}}{\pi^{1+\epsilon}}\bigg[J_0(\bm{q},\bm{k},z)+\sum_{i=1,2}J_1(\bm{q},\bm{k},\bm{l}_i,z)\bigg]=\tfrac{\Gamma^2(\epsilon)\Gamma(1-\epsilon)}{\Gamma(2\epsilon)}\bigg[(\bm{k}^2)^\epsilon[z^{2\epsilon}+(1-z)^{2\epsilon}-1]\\&\qquad\,\,\,+\frac{1}{4}\left\{(\bm{l}_1^2)^\epsilon-[(\bm{l}_1-z\bm{k})^2]^\epsilon+[(\bm{l}_1-\bm{k})^2]^\epsilon-[(\bm{l}_1-(1-z)\bm{k})^2]^\epsilon+\{\bm{l}_1\leftrightarrow\bm{l}_2\}\right\}\bigg].
\end{aligned}
\end{equation}
With the help of the following integrals
\begin{equation}
\begin{aligned}
&\int_0^1\rmd z\frac{(z^{2\epsilon}+(1-z)^{2\epsilon}-1)}{z(1-z)}=-2(\gamma_E+\psi(2\epsilon))=\frac{1}{\epsilon}-\frac{2\pi^2}{3}\epsilon+{\cal O}(\epsilon^2);\\
&\int_0^1\rmd z(z^{2\epsilon}+(1-z)^{2\epsilon}-1)=-1+\frac{2}{1+2\epsilon}=1-4\epsilon+{\cal O}(\epsilon^2);\\
&\int_0^1\rmd z\,z(1-z)[z^{2\epsilon}+(1-z)^{2\epsilon}-1]=-\frac{1}{6}+\frac{1}{1+\epsilon}-\frac{2}{3+2\epsilon}=\frac{1}{6}-\frac{5}{9}\epsilon+{\cal O}(\epsilon^2);\\
&\mu^{-2\epsilon}K_2(\bm{a},\bm{b})=\int_0^1\frac{\rmd z}{z(1-z)}\left[\left[\tfrac{((1-z)(\bm{a}-\bm{b})+z\bm{a})^2}{\mu^2}\right]^\epsilon-(1-z)^{2\epsilon}\left[\tfrac{(\bm{a}-\bm{b})^2}{\mu^2}\right]^\epsilon-z^{2\epsilon}\left[\tfrac{\bm{a}^2}{\mu^2}\right]\right]\\
&\quad=-\frac{1}{\epsilon}-\frac{1}{2}\left[\ln\frac{\bm{a}^2}{\mu^2}+\ln\frac{(\bm{a}-\bm{b})^2}{\mu^2}\right]+\epsilon\left[4\psi'(1)-\frac{1}{2}\ln\frac{\bm{a}^2}{\mu^2}\ln\frac{(\bm{a}-\bm{b})^2}{\mu^2}-\vartheta^2\right]+{\cal O}(\epsilon^2),
\end{aligned}
\end{equation}
we can write
\begin{align}
&\hspace{-1cm}\mathfrak{K}_1\equiv\int_0^1\rmd z P_{gg}(z,\epsilon)\textfrak{s:}_1=2C_a\left[\tfrac{\Gamma^2(\epsilon)\Gamma(1-\epsilon)}{\Gamma(2\epsilon)}\right]\Bigg[(\bm{k}^2)^\epsilon\left[\tfrac{1}{\epsilon}-\tfrac{11}{6}+\left[\tfrac{67}{9}-\tfrac{2\pi^2}{3}\right]\epsilon\right]\\&\hspace{-1cm}+\frac{1}{4}\bigg\{[(\bm{l}_1^2)^\epsilon+[(\bm{l}_1-\bm{k})^2]^\epsilon]\left[-\tfrac{1}{\epsilon}-\tfrac{11}{6}+\tfrac{2\pi^2}{3}\epsilon\right]-2K_2(\bm{l}_1,\bm{k})+4K_3(\bm{l}_1,\bm{k})-2K_4(\bm{l}_1,\bm{k})+\{\bm{l}_1\leftrightarrow\bm{l}_2\}\bigg\}\Bigg].\notag
\end{align}
The terms proportional to $J_2(\bm{q},\bm{k},\bm{l}_1,\bm{l}_2)$ do not involve any $z$ dependence. Since $P_{gg}^{(1)}(z,\epsilon)$ is finite as $z\to 1$, enforcing the diffractive mass cutoff results in 
\begin{align}
&\lim_{\hat{M}_{X,\,{\rm max}}^2\to \infty}\int_0^1\rmd z\,2P_{gg}^{(1)}(z,\epsilon)\Theta\left[\hat{M}_{X,\,{\rm max}}^2-\frac{\bm{\Delta}^2}{z(1-z)}\right]=\lim_{\hat{M}_{X,\,{\rm max}}^2\to \infty}\int_{\frac{\bm{q}^2}{\hat{M}_{X,\,{\rm max}}^2}}^1\rmd z\,2P_{gg}^{(1)}(z,\epsilon)\notag\\&=2C_a\left[2\ln\frac{\hat{M}_{X,\,{\rm max}}^2}{\bm{q}^2}-\frac{11}{6}\right],
\end{align}
where we have discarded power terms in the cutoff. As a second step, we will need the following results to evaluate the momentum integration
\begin{equation}
\begin{aligned}
{\cal I}(\bm{a},\bm{b},M^2)&=\int\frac{\rmd^{2+2\epsilon}\bm{q}}{\pi^{1+\epsilon}}\frac{(\bm{a}-\bm{b})^2}{(\bm{q}-\bm{a})^2(\bm{q}-\bm{b})^2}\ln\frac{M^2}{\bm{q}^2}\\&=\frac{\Gamma^2(\epsilon)\Gamma(1-\epsilon)}{\Gamma(2\epsilon)}\bigg\{\epsilon K_1(\bm{a},\bm{b})+\ln\frac{M^2}{(\bm{a}-\bm{b})^2}[(\bm{a}-\bm{b})^2]^\epsilon\bigg\};\\\mu^{-2\epsilon}K_1(\bm{a},\bm{b})&=\frac{1}{2}\left[\frac{(\bm{a}-\bm{b})^2}{\mu^2}\right]^\epsilon\bigg\{\frac{1}{\epsilon^2}\left[2-\left[\frac{\bm{a}^2}{(\bm{a}-\bm{b})^2}\right]^\epsilon-\left[\frac{\bm{b}^2}{(\bm{a}-\bm{b})^2}\right]^\epsilon\right]\\&\qquad\qquad\qquad\qquad+\ln\left[\frac{\bm{a}^2}{(\bm{a}-\bm{b})^2}\right]\ln\left[\frac{\bm{b}^2}{(\bm{a}-\bm{b})^2}\right]+4\epsilon\psi''(1)+{\cal O}(\epsilon^2)\bigg\}.
\end{aligned}
\end{equation}
Defining
\begin{equation}
{\cal J}(\bm{a},\bm{b})=2C_a\frac{\Gamma^2(\epsilon)\Gamma(1-\epsilon)}{\Gamma(2\epsilon)}\left\{\left[-\frac{11}{6}+2\ln\frac{\hat{M}_{X,\,{\rm max}}^2}{(\bm{a}-\bm{b})^2}\right][(\bm{a}-\bm{b})^2]^\epsilon+2\epsilon K_1(\bm{a},\bm{b})\right\},
\end{equation}
we obtain
\begin{equation}
\begin{aligned}
\mathfrak{K}_2&\equiv\int_0^1\rmd z\int\frac{\rmd^{2+2\epsilon}\bm{q}}{\pi^{1+\epsilon}}\Theta\left(\hat{M}_{X,\,{\rm max}}^2-\frac{\bm{\Delta}^2}{z(1-z)}\right)[2P_{gg}^{(1)}(z,\epsilon)]J_2(\bm{q},\bm{k},\bm{l}_1,\bm{l}_2)\\&=\frac{1}{4}\{{\cal J}(\bm{k},\bm{l}_1)+{\cal J}(\bm{k},\bm{k}-\bm{l}_1)+\{\bm{l}_1\leftrightarrow\bm{l}_2\}\}\\&-\frac{1}{8}\{{\cal J}(\bm{l}_1,\bm{l}_2)+{\cal J}(\bm{k}-\bm{l}_1)+{\cal J}(\bm{l}_1,\bm{k}-\bm{l}_2)+{\cal J}(\bm{k}-\bm{l}_1,\bm{k}-\bm{l}_2)\},
\end{aligned}
\end{equation}
and finally
\begin{equation}
h^{(1)}_{r,\,ggg}=\frac{h_g^{(0)}}{2}\frac{\alpha_{s,\epsilon}}{2\pi}\frac{1}{\mu^{2\epsilon}\Gamma(1-\epsilon)}(\mathfrak{K}_1+\mathfrak{K}_2).
\end{equation}
Expanding around $\epsilon=0$ we find
\begin{equation}\label{polanski}
\begin{aligned}
h^{(1)}_{r,\,q\bar{q}g}&=v^{(0)}\frac{\alpha_s}{2\pi}(2n_f)\frac{1}{6\epsilon}[C_a^2+2C_f^2]+{\cal O}(\epsilon^0);\\h^{(1)}_{r,\,ggg}&=h_g^{(0)}\frac{\alpha_{s,\epsilon}}{2\pi}\frac{2C_a}{\epsilon}\left[\frac{1}{\epsilon}+\ln\frac{\bm{k}^2}{\mu^2}-\frac{11}{3}\cdot\frac{3}{4}+\ln\frac{\hat{M}_{X,\,{\rm max}}^2}{\bm{k}^2}\right]+{\cal O}(\epsilon^0).
\end{aligned}
\end{equation}
It is now easy to see that the pole terms \eqref{polanski} cancel against those of the collinear counterterm \eqref{polchinski}, the UV counterterm \eqref{uvc} and the virtual corrections \eqref{expansion}.

\end{document}